\RequirePackage{etoolbox}
\csdef{input@path}{%
 {sty/}
 {FinalPlots/}
 {RaffaFigs/}
}%

\documentclass[ba,preprint]{imsart}

\usepackage{amsthm}
\usepackage{amsmath}
\usepackage{natbib}
\usepackage[colorlinks,citecolor=blue,urlcolor=blue,filecolor=blue,backref=page]{hyperref}
\usepackage{graphicx}

\usepackage{multirow}
\usepackage{caption}
\usepackage{mathrsfs, latexsym, amsfonts, amssymb, bm}
\usepackage{upgreek} 
\usepackage{threeparttable}
\usepackage{xcolor}
\usepackage{algorithm2e}
\usepackage{framed}
\usepackage{subfig}
\usepackage{bbm}
\usepackage{epstopdf}
\usepackage{appendix}
\usepackage{wela}
\usepackage{multirow, array} 
\newcolumntype{C}[1]{>{\centering\let\newline\\\arraybackslash\hspace{0pt}}m{#1}}

\startlocaldefs
\newcommand{\ind}{\mathbbm{1} }
\newcommand\bbE{\mathbb{E}} 
\newcommand\bbP{\mathbb{P}} 

\newcommand{\bc}{{\bf c}}
\newcommand{\bD}{{\bf D}}

\newcommand{\bL}{{\bf L}}
\newcommand{\bP}{{\bf P}}
\newcommand{\bt}{{\bf t}}
\newcommand{\bU}{{\bf U}}

\newcommand{\bv}{{\bf v}}
\newcommand{\bw}{{\bf w}}
\newcommand{\bx}{{\bf x}}
\newcommand{\bX}{{\bf X}}
\newcommand{\bY}{{\bf Y}}
\newcommand{\by}{{\bf y}}
\newcommand{\bZ}{{\bf Z}}
\newcommand{\bz}{{\bf z}}
\newcommand{\bbeta}{\boldsymbol \beta}
\newcommand{\bepsilon}{\boldsymbol{\epsilon}}
\newcommand{\btheta}{\boldsymbol \theta}
\newcommand{\bvartheta}{\boldsymbol \vartheta}
\newcommand{\bmu}{\boldsymbol \mu}

\newcommand{\bxi}{\boldsymbol{\xi}}
\newcommand{\brho}{\boldsymbol \rho}
\newcommand{\bvarrho}{\boldsymbol \varrho}
\newcommand{\bSigma}{\boldsymbol{\Sigma}}

\newcommand{\btau}{\boldsymbol \tau}
\newcommand{\bphi}{\boldsymbol{\phi}}
\newcommand{\bpsi}{\boldsymbol{\psi}}

\newcommand{\scJ}{\mathcal{J}}

\newcommand\Bern{{\rm Bern}} 
\newcommand\Beta{{\rm Beta}} 
\newcommand{\diag}{{\rm diag}} 
\newcommand{\Gam}{{\rm Gamma}} 
\newcommand{\IW}{{\rm IW}} 
\newcommand{\MN}{{\rm MN}} 
\newcommand\N{{\rm N}} 
\newcommand{\lN}{{\rm logN}} 
\newcommand\U{{\rm U}} 

\newcommand{\ps}[1]{\left[#1\right]} 
\newcommand{\pv}[1]{\left\vert#1\right\vert} 

\newcommand{\iidsim}{\stackrel {{\scriptstyle i.i.d.}}{\sim}} 
\newcommand{\indsim}{\stackrel {{\scriptstyle ind}}{\sim}} 
\makeatletter
\newcommand*{\transpose}{%
    {\mathpalette\@transpose{}}%
}
\newcommand*{\@transpose}[2]{%
    \raisebox{\depth}{$\m@th#1\intercal$}%
}
\makeatother

\usepackage[normalem]{ulem}
\makeatother
\endlocaldefs

\begin{document}

\begin{frontmatter}
\title{{Colombian Women's Life Patterns: A Multivariate Density Regression Approach}}
\runtitle{Multivariate Density Regression}

\begin{aug}
\author{\fnms{Sara} \snm{Wade}\thanksref{addr1}\ead[label=e1]{sara.wade@ed.ac.uk}},
\author{\fnms{Raffaella} \snm{Piccarreta}\thanksref{addr2}\ead[label=e2]{raffaella.piccarreta@unibocconi.it}},
\author{\fnms{Andrea} \snm{Cremaschi}\thanksref{addr4}\ead[label=e3]{andrea.cremaschi@yale-nus.edu.sg}},
\and
\author{\fnms{Isadora} \snm{Antoniano-Villalobos}\thanksref{addr5, addr3}
    \ead[label=e4]{isadora.antoniano@unive.it}} 

\runauthor{S. Wade et al.}

\address[addr1]{School of Mathematics, University of Edinburgh, Edinburgh 
    , UK 
    \printead{e1} 
}

\address[addr2]{Department of Decision Sciences, Dondena Research Centre, Bocconi Institute for Data Science and Analytics (BIDSA), Bocconi University, 
    Milan, Italy
    \printead{e2}
}

\address[addr4]{Yale-NUS College, Singapore
     \printead{e3}
}

\address[addr5]{Department of Environmental Sciences, Informatics and Statistics, Ca' Foscari University of Venice and BIDSA, Italy
    \printead{e4}
}



\end{aug}

\begin{abstract}
Women in Colombia face difficulties related to the patriarchal traits of their societies and well-known conflict afflicting the country since 1948. 
In this critical context, our aim is to study the relationship between baseline socio-demographic factors and variables associated to fertility, partnership patterns, and work activity. To best exploit the explanatory structure, we propose a Bayesian multivariate density regression model, which can accommodate mixed responses with censored, constrained, and binary traits. The flexible nature of the models allows for nonlinear regression functions and non-standard features in the errors, such as asymmetry or multi-modality. The model has interpretable covariate-dependent weights constructed through normalization, allowing for combinations of categorical and continuous covariates.  Computational difficulties for inference are overcome through an adaptive truncation algorithm combining adaptive Metropolis-Hastings and sequential Monte Carlo to create a sequence of automatically truncated posterior mixtures. For our study on Colombian women's life patterns, a variety of quantities are visualised and described, and in particular, our findings highlight the detrimental impact of family violence on women's choices and behaviors.
\end{abstract}



\end{frontmatter}

\section{Introduction}
Colombian women face difficulties that are quite typical in Latin American countries, particularly related to the patriarchal traits of their society. 
Nonetheless, the welfare of Colombian women is possibly more critical due to the conflict between state military forces, paramilitaries, and guerrilla groups that has afflicted the country since 1948. 
As underlined by \citeauthor{Gimenez_etal_2015}, dramatic subnational inequalities exist in every indicator, especially within low-income, low-education, and rural populations. Reinforcing constraints, such as ``limited and gender-unequal economic opportunities, exclusion from quality endowments among marginalized
populations, and social norms and gender roles that relegate unpaid care work to women
and tolerate violence against them (emotional, physical and sexual), affect young women's
choices and actions with respect to life plans and fertility decisions''\citep[][p. 5]{Gimenez_etal_2015}. In particular, despite significant progress since 2000, teenage pregnancy rates in Colombia are still very high. The majority of teenage pregnancies remain unplanned, signaling a lack of opportunity and agency for young girls.  Different studies discuss the detrimental effects of teenage pregnancy \cite[see e.g.,][]{Gimenez_etal_2015,Azevedo_2012} and its socio-demographic drivers, such as poverty, low levels of education, and living in rural areas. 
In such a critical context, we are interested in studying women's life events, focusing on the interplay between sexual initiation (debut), fertility, partnership, and participation in the labor market. 
Thus, rather than focusing on a specific life event, as in previous relevant studies \citep[e.g.][]{Gimenez_etal_2015,Azevedo_2012,martinez2017sexual}, we adopt a broader perspective, considering a collection of events describing transition to adulthood and their relation with a set of structural baseline characteristics of the women's environment and family. Besides some of the well known relevant factors, such as cohort, region, and area (urban or rural) of residence, we also study whether a violent family context contributes to shape transition to adulthood and possibly impairs women's agency. 

To this purpose, we analyze data arising from the survey conducted in Colombia in 2010 as a part of the Demographic and Health Survey Program (DHS,  \url{https://www.dhsprogram.com}).
The data are cross-sectional, thus, no follow-up information on the life events of interest is recorded. Specifically, information is available on the age when the considered focal events (sexual debut, marriage or cohabitation, motherhood) were experienced for the first time, whereas work information concerns only the employment status of the woman (working or not) at the moment of the interview. 
Thus, we jointly analyze response variables with different levels of measurements (times at event and binary variables). 
Additionally, the focal events may not have been experienced (right-censoring) and are subject to 
constraints, e.g. motherhood can only occur after sexual debut. 
Furthermore, the available set of baseline explanatory variables is limited, so that heterogeneity may be present which would not be properly captured by a parametric model. This encourages the use of a flexible model to best exploit the explanatory structure without imposing possibly penalizing constraints. Additionally, such a model can encompass competing sociological theories which may be relevant in different subpopulations.

The data present various features that challenge existing parametric and semiparametric models \citep[e.g.][]{Korsgaard2003, jaraSemiparCensored2010, hanson2004bayesian, kottas2001bayesian}. First, some women postpone the events to relatively late in life, which induces right-skewed distributions. Also, the joint relationships between the age-at-event variables show different patterns, with gaps of various lengths between events. Moreover, these behaviors change depending on the covariates. 
Modeling such dependence structure is an ambitious task, requiring a model that 
allows for i) non-linear response curves, ii) non-normal
distributions whose features may change with the covariates, iii) multivariate response and covariates of mixed nature, and iv) censoring and constraints of the responses. To the best of our knowledge, a model that can simultaneously deal with these issues does not exist. 

We propose a Bayesian multivariate density regression model that extends the univariate model of \cite{AWW2014} to the case of multiple mixed-type responses with censoring and constraints. This approach is promising for our data, due to its ability to capture  their peculiar features. 
Our infinite mixture model has interpretable covariate-dependent weights constructed through normalization, allowing for combinations of categorical and numerical covariates. In addition, the multivariate approach permits us  to study the joint relationship between the response variables, for example by considering one response conditioned on the others. 
With data on over 10,000 women and a multivariate response and covariate, the Markov chain Monte Carlo (MCMC) algorithm originally proposed for the univariate model becomes unsuitable. We therefore propose an algorithm for posterior inference that extends the adaptive truncation scheme of
\cite{Griffin16}.

The paper is structured as follows. Section \ref{sec:data} describes the data. 
The model and posterior simulation algorithm are presented in Sections \ref{sec:BNPdensityregression} and \ref{sec:AdaptiveTrunc}, respectively. The
results for the data on Colombian women are analyzed in Section \ref{sec:Application}. Section \ref{sec:Conclusions} summarizes and concludes. In addition, the Supplementary Material (SM) includes derivations and details for predictive inference, as well as additional results for both the simulated data example and the case study.

\section{The data}\label{sec:data}
The DHS Program collects and disseminates data on random samples of households selected from random clusters from a national sampling frame\footnote{Note that the data arise from a complex survey design, and thus have associated weights with a complicated structure. Moreover, additional post-stratification is carried out to adjust the weights for various factors, such as the total number of women interviewed in each municipality, non-response, etc. \citep[see][for full details on the survey design and weighting scheme for the Colombian survey]{Profamilia_2011_DHS}. A discussion on the suitability of accounting for such sampling weights is offered in the SM.}. 
The 2010 survey in Colombia was conducted by the Profamilia association, and we refer to the final report for a detailed description of its features \citep{Profamilia_2011_DHS}. Since all women of childbearing potential (i.e. aged 13-49) in the same household were interviewed, we randomly select at most one case from each household to avoid unwanted dependencies.

To describe the characteristics of fertility and partnership patterns, we consider the discrete variables recording the ages at \emph{Sexual Debut} ($Z_1$),  at \emph{Union} (first marriage or cohabitation, $Z_2$), and at \emph{First Child} ($Z_3$). \emph{Work Status} is recorded as a binary variable ($Z_4$) indicating whether the respondent worked in the 12 months before the interview. We exclude women who gave inconsistent information, namely, those who report the birth of the first child as preceding the first sexual intercourse, and those who report union with a partner but for whom sexual intercourse never occurred.
We also filter out  women who experienced sexual violence or were forced to have sex in exchange for money, since their  union and childbearing choices may be related to the experienced violence. By the same reasoning, we  remove women who were forced to use contraceptive methods. Thus, we attempt to focus as much as possible on life choices and plans rather than on events imposed by circumstances, even if the latter may be unknown and unmeasured, so that the observed events may not necessarily reflect choices.

We focus on the relationship between the responses and some baseline socio-de\-mo\-gra\-phic factors. First, we consider the woman's \emph{Age} (in years) at the moment of interview. We focus on women aged 15 or more, as most younger women had not yet experienced any event at the time of the survey. 
Next, we include the type of \emph{Area}  (urban or  rural) where the respondent lives, as well as her \emph{Region}. Following the partition used in the DHS dataset \citep[as well as in the report by][]{Profamilia_2011_DHS} we consider the five regions Atlantica, Oriental, Central, Pacifica, and Territorios Nacionales\footnote{The former region ``Territorios Nacionales'' includes the two more easterly regions, Orinoquia and Amazonia, created in 1991.}; the capital Bogota, located in the Oriental region, is treated as a sixth region because of its peculiar features in terms of social and economic development. A map of Colombia and  the considered regions is reported in the SM.
Since information is only available on the current region of residence and on the age when respondents moved there, we limit attention to women who were raised in the current region at least from the age of 6, to properly account for regional effects. Moreover, to assess the respondent's well-being in her original family, we consider whether in her childhood she was disciplined using \emph{Physical Punishment} (spanking, hitting, pushing, throwing water), and whether she was exposed to \emph{Parental Domestic Violence} and ever witnessed her father beating her mother. 
All respondents missing at least one explanatory or response variable are excluded from the dataset. 

Even if the DHS dataset is very rich, including other covariates is not straightforward. Most of the variables refer to the moment of interview, and thus cannot be considered as antecedents of the focal events. For example, although it would be interesting to include information regarding education and wealth, only the highest level of education attained 
and the wellness of the respondent's family at the moment of interview are available. 
Another relevant aspect that could be taken into account concerns women's ethnicity. However, most (about 80\%) of the interviewed women  do not recognize themselves as part of an ethnic minority. Furthermore, those who do belong to a heterogeneous variety of ethnic groups, none of which are sufficiently represented in the sample. We therefore exclude ethnic minorities from our study.

\begin{table}
	\centering
	\begin{tabular}{c|cc|cc|cc}
		\multicolumn{1}{c}{}
		& \multicolumn{2}{c}{\emph{Sexual Debut}} &\multicolumn{2}{c}{\emph{Union}} &\multicolumn{2}{c}{\emph{First Child}}\\ \hline
		\emph{Age} & Censored & Observed & Censored &Observed & Censored &Observed \\ \hline
		15--19 & 1,144  & 1,053& 1,818 & 379 & 1,837 & 360 \\
		20--29& 238 & 3,475 & 1,358 & 2,355 & 1,323 &2,390 \\
		30--39 & 51 & 2,597 & 378 & 2,270 & 326 &2,322 \\
		40--49 & 55 & 2,127 & 281 & 1,901 & 216 &1,966 \\\hline
		& 1,488 & 9,252 & 3,835 & 6,905 &3,702 & 7,038
	\end{tabular}
	\caption{Cross-tabulation of age groups and censored data.}
	\label{tbl:censoring}
\end{table}

\begin{figure}
	\begin{center}
		\includegraphics[width=0.9\textwidth]{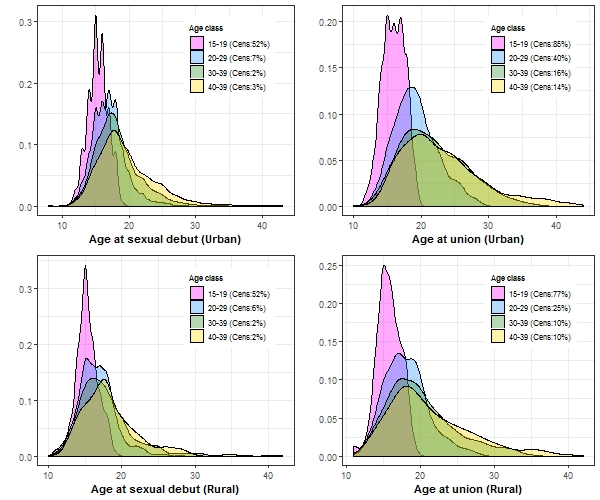} 
		\caption[]{Kernel density estimates of (non-censored) ages at events conditioned to $Age$ (at interview, in groups) and area of residence (Urban or Rural).}
		\label{descr_densities}
	\end{center}
\end{figure}

\begin{figure}
	\begin{center}
		\includegraphics[width=0.99\textwidth]{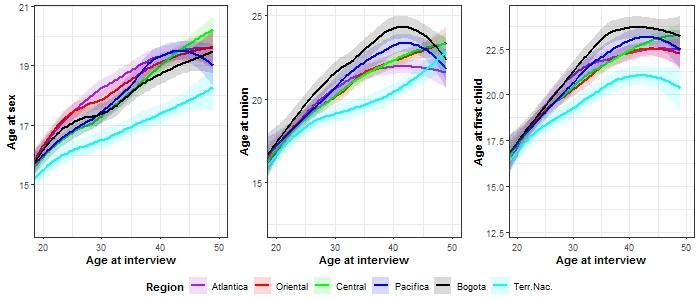} 
		\caption[]{Relation (smoothed regression) between (non-censored) ages at events and $Age$.}
		\label{descr_regress}
	\end{center}
\end{figure}

Our final dataset consists of $n=10740$ women.
Table \ref{tbl:censoring} reports a summary of the number of censored cases for the first three response variables within age groups.
Figures \ref{descr_densities} and \ref{descr_regress} offer some insights about the distinctive features of the densities of the ages at events and of their relationships (smoothed regressions) with the $Age$ at interview across different covariates values. Even if the displayed results only relate to non-censored cases, they emphasize the difficulties implied by modeling the dependence structures. To achieve such an ambitious task, in the next section we propose a model that allows for non-linear response curves (see e.g. Figure \ref{descr_regress}), non-normal distributions whose features may change with the covariates (see e.g. Figure \ref{descr_densities}), multivariate response and covariates of mixed nature, and censoring of the responses\footnote{Please note that we refer specifically to \textit{random} (or non-informative) censoring, i.e. to the case when each subject has a censoring time that is statistically independent of the age at a given response event.}.

\section{Bayesian nonparametric density regression}\label{sec:BNPdensityregression}
We develop a Bayesian nonparametric mixture model that can capture the relationship between $n$ conditionally independent $d$-dimensional response vectors, $\bZ_i$, and a vector  $\bx_i^*$ of predictors. To simplify  notation, whenever possible we drop the sub-index $i$ indicating individual observations. The predictors $\bx^*=(x_1, \ldots,x_p,x^*_{p+1},\ldots,x^*_{q^*})$ may be of mixed nature. Without loss of generality, we assume that the first $p$ are numerical while the rest are categorical. As  is common in regression models, we expand the categorical predictors with binary dummy variables and let $\bx=(x_1, \ldots,x_p,x_{p+1},\ldots,x_{q})$, where $q=p+\sum_{k=p+1}^{q^*} (R_k-1)$ and $R_k$ denotes the number of categories of $x^*_k$.
The observed responses are also of mixed nature. For example,
in our application, we consider two types of responses: three positive integer-valued variables with possible censoring and constraints, representing the ages at events, and one binary variable indicating work status.  
In this case, we refer to the density of the mixed response $\bZ=(Z_1,\ldots,Z_d)$ with respect to the appropriate measure, e.g. Lebsegue or counting measure, for each response type. 

To frame our model within existing literature, we review some related contributions.
Bayesian nonparametric mixture models \citep{lo84} are useful tools for density estimation, due to their attractive balance between flexibility and smoothness and ability to recover a wide range of densities \citep[][Chapter 5]{GhoshRama}. Further developments for conditional density estimation, also known as density regression, can be found in the pioneering works of \cite{MEW} and \cite{Mac1}. Extensions of the former for categorical responses can be found in \cite{Shahbaba}. We focus on the latter work of \cite{Mac1}, which  extends the Bayesian nonparametric mixture model by allowing the mixing measure to depend on the covariates. This yields flexible density regression, where the entire density and not only the mean is regressed on the covariates. Several approaches exist in literature to specify the covariate-dependent mixing measure, but it is not clear how to choose between them. Examples include single-$p$ dependent Dirichlet processes \citep{MacTR, DeIorio}, with covariate-dependent component parameters but single weights. Alternatively, numerous proposals have been introduced for covariate-dependent weights \citep[to name a few]{GS06, Dun1, Rod}. In terms of the support and consistency properties of the model, the distinction is irrelevant \citep[see e.g.][]{Barrientos2012,Barrientos2017, Pati2013}. However, the analysis of \cite{WadeWalkerPetrone} suggests that prediction issues may arise at moderate sample sizes for single weights mixtures with linear predictors, when the true relation between response and covariates is non-linear. Given the complex nature of the application that we are considering and the lack of theory justifying transformations that would simplify the shape of the relation between the variables, we have decided to consider a linear predictor approach. Therefore, in this work, we build on the construction of the covariate-dependent weights developed by \cite{AWW2014}, which allows for combinations of continuous and discrete covariates and favors interpretability.  

We require extending the model to multivariate responses of mixed type with possible censoring and constraints. 
An appealing approach for this relies on a latent Gaussian representation, which provides a simple construction for dependence of the multivariate mixed-type data through the full covariance matrix of the latent Gaussian variables. Moreover, Bayesian inference can be carried out through Gibbs sampling and data augmentation techniques. A Bayesian parametric model based on this idea was proposed by \cite{Korsgaard2003} for multivariate data combining Gaussian, right-censored Gaussian, ordinal, and binary traits. 
To increase model flexibility, Bayesian nonparametric versions were proposed by \cite{DeYoreo2} for mixed ordinal and nominal data, by \cite{DeYoreo} for multivariate ordinal regression \citep[see][for a dynamic extension]{DeYoreo3} and by \cite{PapaRichBest} for mixed-type spatial data in an epidemiological context. Due to the increased flexibility of nonparametric mixtures, the cut-offs used to define the discrete data from the latent Gaussian variables can be fixed and not estimated or inferred. Moreover, \cite{Canale} show that Bayesian nonparametric mixtures for discrete data (specifically counts) based on latent Gaussian variables can approximate and consistently estimate a wider range of distributions than mixtures based on discrete distributions, e.g. Poisson or multinomial (see also \cite{KMQ2005, BaoHanson} in the context of multivariate ordinal data and \cite{Jara2007} for binary response regression). 
Another relevant extension is the Bayesian semiparametric model of \cite{jaraSemiparCensored2010} for multivariate doubly-censored data indicating time to event, based on a log transformation  linking the observed responses to the latent Gaussian variables. When modeling time-to-event data, the log transformation is more appropriate than others, notably truncation. This allows recovery of the underlying structure with fewer and more interpretable components with possibly heavy right tails. Recently, \cite{norets2018adaptive} demonstrated that optimal adaptive estimation of mixed discrete-continuous distributions can be achieved via the latent Gaussian mixture approach.

We combine some of these ideas to build a model which can deal with the challenges presented by the data. We adopt the latent Gaussian approach, associating to each response variable $Z_\ell$ a latent real-valued $Y_\ell$. 
Specifically, an observed value $z_\ell$ of the response $Z_\ell$  is linked to the realization $\by=(y_1,\ldots,y_d)$ of the latent $\bY=(Y_1,\ldots,Y_d)$, through a function $h_\ell$
whose characteristics depend on the nature of the observable.
Examples of transformations for different response types include:
\begin{align*}
z_\ell = h_\ell( \by,\bx)= & y_\ell, \quad \text{for } z_\ell \in \mathbb{R}, \\
z_\ell = h_\ell( \by,\bx)= & \lfloor \exp(y_\ell) \rfloor, \quad \text{for } z_\ell \in \mathbb{N}, \\
z_\ell = h_\ell( \by,\bx)=& \sum_{a=1}^{A_\ell-1} \ind_{[\alpha_{\ell,a},\infty)} (y_\ell), \quad \text{for } z_\ell \in \lbrace 0,1 ,2,\ldots,A_\ell -1\rbrace,
\end{align*}
where $\lfloor \cdot \rfloor$ denotes the floor function, and $\ind_B(y)$ denotes the indicator function taking the value one when $y\in B$. Note that the last case considers an ordinal response with $A_\ell$ categories and fixed cutoffs of $\alpha_{\ell,1}<\ldots<\alpha_{\ell,A_\ell-1}$. In these examples, the functions $h_\ell$ do not depend on $\bx$ or $y_{\ell'}$ for $\ell'\neq\ell$, but they may, for example when accounting for censored or constrained responses, as for the case study described in Section \ref{sec:Application}. 

The basic building block for our model is the multivariate multiple linear regression:
\begin{equation*}
\bY|\bx, \bbeta, \bSigma\stackrel{ind}{\sim} \N_d(\by|\bx\bbeta,\bSigma), 
\end{equation*}
where $\bbeta$ is a $(q+1) \times d$ matrix of regression parameters and $\bSigma$ is a $d\times d$ covariance matrix. 
Slightly abusing notation, $\bx = (1,x_1, \ldots, x_{q})$ denotes the vector of observed covariate values extended by a unitary entry.   
As previously discussed, this parametric model is not flexible enough to capture the complex dependence structures contained in the data.
We therefore extend the nonparametric density regression framework introduced by \cite{AWW2014} to model the $\mathbb R^d$-valued latent variable $\bY$: 
\begin{equation}\label{mv_AWW}
f_{\bP_\bx}(\by| \bx)= \sum_{j=1}^\infty w_j(\bx) \N_d(\by|\bx\bbeta_j,\bSigma_j),\quad \text{with}\quad
w_j(\bx) =\frac{w_j\, g(\bx|\bpsi_j)}{\sum\limits_{j'=1}^\infty w_{j'}\, g(\bx|\bpsi_{j'})}.
\end{equation}
This model results from considering a mixture
\begin{equation*}
f_{\bP_\bx}(\by| \bx)= \int \N_d(\by|\bx\bbeta,\bSigma)\text{d}\bP_\bx(\btheta), 
\end{equation*}
where $\btheta = (\bbeta, \bSigma)$ and a nonparametric prior is assigned to the set of covariate-dependent mixing measures $\bP_\bx$, which places mass one on the set of discrete probability measures:
\begin{equation*}
\bP_\bx = \sum_{j=1}^\infty w_j(\bx)\, \updelta_{\btheta_j}. 
\end{equation*}
Here, $\updelta_\theta$ denotes the Dirac-delta function with unit mass at $\theta$. For computational purposes and to ensure convergence of the normalizing constant in $w_j(\bx)$, it is convenient to adopt a stick-breaking representation for the weights, setting
$w_1=v_1$ and $w_j=v_j \prod_{j' < j} (1-v_{j'})$, for $j >1$, where $v_j \indsim \Beta(\zeta_{j,1}, \zeta_{j,2})$. The parameters of the local linear regression components, $\btheta_j$, and of the covariate-dependent weights, $\bpsi_j$, are assumed to be independent and identically distributed according to a base measure $\bP_0$ and independent of the weights. Together with the functions $h_\ell$ linking the latent variables with the responses, this defines the likelihood structure for the observed data.

In this model, the regression parameters $\bbeta_j$ and $\bSigma_j$ capture the local linear relation between the latent response and covariates, with normal errors; whereas the $\bpsi_j$ determine, through $g$, how the influence of each local component to the overall model changes across the covariate space. 
This deals with situations when the stochastic relation between $\by$ and $\bx$ is too complicated to be captured by a single parametric model. It can also be used when the population is assumed to be constituted by an unknown number of (covariate-dependent) groups such that, within each group, a linear regression model provides a good description of the data. While identifiability issues may prevent the individuation of such groups, this intuition can help in understanding the elements of the model. 

Note that the Bayesian nonparametric model for the joint density of $\by$ and $\bx$ introduced by \cite{MEW} for density regression, taking the form 
\begin{equation}\label{eq:joint}
f_{\bP}(\by, \bx)=  \sum_{j=1}^\infty w_j \, g(\bx|\bpsi_j)\, \N_d(\by|\bx\bbeta,\bSigma), 
\text{ with } 
\bP = \sum_{j=1}^\infty w_j \, \updelta_{(\btheta_j,\bpsi_j)}, 
\end{equation}
results in a conditional density coinciding with equation \eqref{mv_AWW}. However, an important difference is that in the joint mixture model, posterior inference for the parameters $(w_j, \btheta_j,\bpsi_j)$ is based on the joint likelihood in \eqref{eq:joint}; whereas, for our model, it is based directly on the conditional likelihood of interest. Furthermore, we emphasize that the converse is not true: our conditional density model in \eqref{mv_AWW} does not imply the joint density model in \eqref{eq:joint}. This can be easily seen by constructing a joint density model as the product of \eqref{mv_AWW} and any, say parametric, marginal density model for $\bx$. This is a valid construction, which nonetheless recovers the joint model in \eqref{eq:joint} only when the marginal has the form:
\begin{equation*}
f_{\bP}(\bx)=  \sum_{j=1}^\infty w_j \, g(\bx|\bpsi_j). 
\end{equation*}
This is an important concept, as it highlights that the form chosen for $g$ does not imply a modeling of the distribution for covariates, which may indeed be fixed. The choice and shape of this kernel, however, defines how the conditional distribution changes as $\bx$ varies (given the parameters $\bpsi$). Thus, it determines the amount of information borrowed when making inference at  unobserved points in the space of covariates. By choosing model \eqref{mv_AWW} we maintain the same natural and interpretable structure for the weights of the joint mixture model, but exploit all the information available in the data to learn about the relation between $\bx$ and $\by$, thus improving the quality of the estimation for the conditional distribution \citep[see][]{wade2014improving}, which is the main focus of our application. Clearly, a practitioner interested also in capturing the structure of the random covariates would require a different approach.

The covariate-dependent weight $w_j(\bx)$ represents the probability that an observation with a covariate value $\bx$ is allocated to the $j$-th regression component. Such probability can be decomposed into the unconditional probability $w_j$ that  the parametric model $j$ fits an individual observation, and the likelihood $g(\bx|\bpsi_j)$ that an individual allocated to the $j$-th component is characterized by a covariate value $\bx$. The $g(\cdot|\bpsi)$ can be defined to accommodate different types of covariates. We adopt a factorizable structure:
\begin{equation*} 
g(\bx|\bpsi) = \prod_{k = 1}^{q} g(x_k|\psi_k),\quad\text{where}\quad
g(x_k|\psi_k) = \left\lbrace \begin{array}{ll} \N (x_k | \mu_{k}, \tau_{k}^{-1})& \text{ for } k =1,\ldots,p,  \\
\Bern(x_{k}|\rho_{k})& \text{ for } k = p+1, \ldots, q, \\
\end{array} \right. 
\end{equation*}
with $\psi_k = (\mu_{k},\tau_{k})$ for $k =1,\ldots,p$, and $\psi_k = \rho_{k}$ for $k = p+1, \ldots, q$. The use of distribution kernels guarantees convergence, for all $\bx$, of the denominator in equation \eqref{mv_AWW}. For the unconditional probability $w_j$, different choices of the stick-breaking parameters $(\zeta_{j,1}, \zeta_{j,2})$ result in different nonparametric priors \cite[see][]{IJ}. For instance, if $(\zeta_{j,1}, \zeta_{j,2})= (1,\zeta)$,
the prior on the weights $w_j$ corresponds to that obtained from a Dirichlet process prior. 
The base measure is chosen as $\bP_0(\bbeta, \bSigma, \bpsi) = \bP_0(\bbeta| \bSigma)\bP_0(\bSigma) \bP_0(\bmu|\btau)\bP_0(\btau) \bP_0(\brho).$
We use the conjugate  matrix-variate Normal-In\-verse Wishart for the regression parameters: $\bP_0(\bbeta| \bSigma) = \MN_{(q+1) \times d}(\bbeta_0, \bU,\bSigma)$, where $\bbeta_0$ is a  $(q+1) \times d$ matrix and $\bU$ is a $(q+1) \times (q+1)$ positive definite matrix; and $\bP_0(\bSigma) = \IW(\bSigma_0, \nu)$,
where $\bSigma_0$ is a $d\times d$ positive definite matrix and $\nu>0$. Notice that the Inverse Wishart assigns prior mass to full covariance matrices. Other prior specifications can be used to allow for other types of covariance structures, e.g. a product of Inverse Gammas for diagonal covariance matrices or a G-Wishart for sparse precision matrices. As for $\bbeta$, we are assuming a structured dependence, allowing for efficient computations through Kronecker products and a reduced number of hyperparameters compared to a full Gaussian distribution. Alternatively, a multivariate Gaussian distribution could be used, assuming independence between columns. To complete the specification of the base measure, we set: 
$\bP_0(\bmu|\btau)  =   \prod_{k=1}^{p}\N \left(\mu_{k}|\mu_{0,k},(u_{k} \cdot \tau_{k})^{-1} \right)$, $\bP_0(\btau) = \prod_{k=1}^{p}\Gam (\tau_{k}| \alpha_{k},\gamma_{k})$, and $\bP_0(\brho)  =  \prod_{k=p+1}^{q} \Beta(\rho_k | \bvarrho_k)$, where $\bvarrho_k = (\varrho_{k,1},\varrho_{k,2})$.

In the next section, we describe an adaptive truncation algorithm allowing posterior inference for our model. The algorithm is general and only requires specific adjustments depending on the $h_\ell$ functions linking the observed responses with their latent counterparts.

\section{Adaptive truncation algorithm}\label{sec:AdaptiveTrunc}
To scale appropriately with the sample size and data dimensions, we devise an algorithm for posterior inference based on a finite truncation of the mixture, where the number of components is allowed to increase adaptively to obtain a good approximation of the infinite-dimensional posterior. 
The truncated latent model with $J$ components is:
\begin{align}\label{eq:trunc}
f_{\bP_\bx^J}(\by| \bx)&= \sum_{j=1}^J w_j^J(\bx) \N_d(\by|\bx \bbeta_j, \bSigma_j),
\end{align}
where the weights follow the re-normalized stick breaking construction: 
\begin{equation}\label{eq:truncweight}
w_j^J(\bx)=\frac{w_j g(\bm x|\bpsi_j)/\sum_{j=1}^{J}w_j}{\sum_{j'=1}^J w_{j'}g(\bm x|\bpsi_{j'})/\sum_{j=1}^{J}w_j}=
\frac{w_j g(\bx|\bpsi_j)}{\sum_{j'=1}^J w_{j'} g(\bx|\bpsi_{j'})}.
\end{equation}
Notice that the normalizing constant  $\sum_{j=1}^{J}w_j$ in \eqref{eq:truncweight} cancels out. To ease notation, we  use $w_j(\bx)$ to denote the truncated covariate-dependent weights, dropping the superscript $J$ when the truncation level is clear. Due to the exponential decay of the weights, for large enough $J$, the truncated model (\ref{eq:trunc}) provides a close approximation to the infinite mixture model. Alternative truncation methods could be considered, notably the popular truncated stick breaking method \citep{IJ} where $v_J=1$.  
However, re-normalized stick-breaking may provide a better finite-dimensional approximation by evenly distributing the remaining mass across components, as opposed to assigning all remaining mass to the last component in  truncated stick-breaking.

The proposed algorithm is based on the adaptive truncation scheme  developed by \cite{Griffin16}, extended for density regression and mixed type responses. It consists of two main steps, namely an  MCMC step for a fixed truncation level, $J_0$, followed by a sequential Monte Carlo (SMC) step used to increase the number of components of the mixture. The first step produces $M$ posterior draws $(\bw_{1:J_0}^m,\btheta_{1:J_0}^m,\bpsi_{1:J_0}^m, \by_{1:n}^m)_{m=1}^M$, which are then used as particles in the SMC step. 
We provide a concise summary below, with software and full details provided through the authors' GitHub repository\footnote{\url{https://github.com/sarawade/BNPDensityRegression_AdaptiveTruncation}} and accompanying documentation. 
\paragraph{MCMC for fixed truncation.}
Since the truncation level $J_0$ is fixed throughout this step, we omit it from the notation, writing $\bw = w_{1:J_0}$, $\btheta=\btheta_{1:J_0}$, and $\bpsi=\bpsi_{1:J_0}$. Similarly, the observed response is denoted by $\bz = (\bz_1,\ldots,\bz_n)$, with $\bz_i = (z_{i,1}, \ldots, z_{i,d})$, and analogously for the covariates $\bx$ and the latent $\by$. 
The approximate posterior given the sample $(\bx,\bz)$ of size $n$, using the truncated likelihood \eqref{eq:trunc}, takes the form: 
\begin{align*} 
\bP_{J_0}^n(\bw, \bpsi, \btheta, \by|\bz, \bx) \propto & \,\bP_{J_0}(\bw,\bpsi,\btheta)\prod_{i=1}^n \sum_{j=1}^{J_0} w_j(\bx_i|\bpsi_j) \N_d(\by_i|\bx_i \bbeta_j,\bSigma_j)   \prod_{\ell=1}^{d} \ind_{\{z_{i,\ell}\}}(h_{i,\ell}),
\end{align*}
where $\bP_{J_0}(\bw,\bpsi,\btheta)$ indicates the restriction of the prior (as detailed in Section \ref{sec:BNPdensityregression}) to the parameters in the truncated space. Dependence $w_j(\bx)=w_j(\bx|\bpsi_j)$ of the weights on the parameters has been made explicit. 
Moreover, the functions $h_{i,\ell} = h_\ell(\by_{i},\bx_i)$ linking the latent variables to the observed responses are tailored to the specific application in Section \ref{sec:Application}.

Due to lack of conjugacy, we resort to a generic Metropolis-within-Gibbs scheme to perform posterior sampling, that updates blocks of parameters adaptively. The adaptive random walk algorithm used here, based on Algorithm 6 in \cite{GS}, adapts the proposal covariance matrix to achieve both a specified average acceptance rate ($a_0=0.234$) and a proposal covariance matrix equal to $2.4^2/\mathfrak{p}$ times the posterior covariance matrix, $\mathfrak{p}$ being the dimension of the parameter block of interest. These criteria have been shown to be optimal in many settings \citep{RGG97, RR01}. In more detail, suppose that we want to sample a block of parameters $\bphi$ of dimension $\mathfrak{p}$ from a distribution with probability density function $Q$. In order to utilise the adaptive random walk algorithm, we first consider a transformation $t(\bphi)$  that has full support on $\mathbb{R}^{\mathfrak{p}}$. 
At each iteration $m$, we propose a new  $\bphi^*$ such that:
\begin{equation}\label{eq:AMHproposal}
\bt^*\equiv t(\bphi^*)= t(\bphi^{m-1})+\bepsilon,\text{  with } \bepsilon\sim \N(0, \bxi^{m-1}),
\end{equation} 
where $ \bxi^{m-1}$ is the adaptive covariance matrix. We accept $\bphi^m = \bphi^*$ with probability 
equal to the minimum between 1 and the ratio: 
\begin{equation}\label{eq:aratio}
a(\bphi^*,\bphi^{m-1})=\frac{Q(\bphi^*)}{Q(\bphi^{m-1})} \frac{\pv{\mathcal{J}_t(\bphi^{m-1})}}{\pv{\mathcal{J}_t(\bphi^*)}},
\end{equation}
with $\pv{\mathcal{J}_t(\bphi)}$ denoting the determinant of the Jacobian of the transformation.

Transformations of $\bbeta_j$, $\bmu_j$, $\btau_j$, $\brho_j$, and $\bv_j$ are straightforward through identity, log, and logit functions. Instead, transformations of $\bSigma_j$ and $\by_i$ are more involved. For each $\bSigma_j$, we consider a vectorization of a decomposition of the matrix, $\bSigma_j = \bL_j\bD_j\bL_j^\transpose$, where $\bL_j$ is a lower triangular matrix with unit entries on the diagonal and $\bD_j$ is a diagonal matrix with positive entries, and we take the log of the diagonal entries. In this case, the proposed $\bSigma_j^*$ can be obtained from  the proposed $\bt^*$ in equation \eqref{eq:AMHproposal} by inverting this transformation. In addition, the determinant of the Jacobian, which is required in the acceptance ratio in \eqref{eq:aratio}, depends only on the diagonal elements $D_{j,\ell,\ell}$ of the matrix $\bD_j$, specifically, $\pv{\scJ_t(\bSigma_j)}=\prod_{\ell=1}^d 1/D_{j,\ell,\ell}^{d+1-\ell}$. 
For each latent vector $\by_i=(y_{i,1},\ldots,y_{i,d})$, the terms $h_\ell(\by_i,\bx_i)=z_{i,\ell}$ define constrained regions for the latent $\by_i$, such that $y_{i,\ell} \in (l_{i,\ell},u_{i,\ell})$, which are provided for the case study in Section \ref{sec:Application}.  
We assume that an appropriate ordering of the responses leads to bounds $(l_{i,\ell},u_{i,\ell})$ that may in general  depend on $y_{i,\ell'}$ for $\ell' < \ell$. 
This allows us to define a sequential logistic transformation $t(y_{i,\ell}; \by_{i,1:\ell-1})$ for $\ell=1,\ldots,d$, based on the bounds $(l_{i,\ell},u_{i,\ell})$. From the proposed $\bt^*$ in equation \eqref{eq:AMHproposal}, the inverse transformation can be applied to obtain the proposed $\by_i^*$, 
sequentially for $\ell=1,\ldots,d$, where the bounds may also be updated sequentially if they depend on $y_{1:(\ell-1)}^*$, e.g. for age at first child in our application.
This ordering also guarantees that the Jacobian matrix is lower triangular, so its determinant is simply the product of the diagonal elements, $\pv{\scJ_t(\by_i)}=\prod_{\ell=1}^d \scJ_{t,\ell,\ell}(y_{i,\ell}; \by_{i,1:\ell-1})$, with $\scJ_{t,\ell,\ell}(y_{i,\ell}; \by_{i,1:\ell-1})=(u_{i,\ell}-l_{i,\ell})/\left[ (y_{i,\ell}-l_{i,\ell})(u_{i,\ell}-y_{i,\ell}) \right]$, for $u_{i,\ell} \in \mathbb{R} , l_{i,\ell} \in \mathbb{R} $.

\paragraph{SMC for adaptive truncation.}
The second stage involves the selection of the truncation level $J$
by sequentially increasing it from the initial level $J_0$. The addition of a new component improves the quality of the approximation to the infinite-dimensional model but increases the computational burden, due to the considerable number of parameters added. Therefore, devising an algorithm that can select the level of truncation parsimoniously is crucial. To achieve this, possible approaches are presented in \cite{norets2017optimal} and \cite{Griffin16}. We focus on the latter, which adaptively increases the number of mixture components via SMC.

The MCMC draws from the previous step are used as the $M$ initial particles in the SMC.
At each iteration of the SMC, a new component is added to the mixture, by sampling the additional set of parameters $(w^{m}_{J + 1}, \bpsi^{m}_{J + 1}, \btheta^{m}_{J + 1})$ from a suitable importance distribution. We sample from the prior $ \Beta(v^{m}_{J+1}) \bP_0(\bpsi^{m}_{J + 1}, \btheta^{m}_{J + 1}),$ 
independently for $m=1,\ldots,M$, 
making use of the recursive stick-breaking relation $w^{m}_{J + 1} = v^{m}_{J + 1}\ps{(1 - v^{m}_J)/v^{m}_J}w^{m}_J$.
The particle weights $\tilde{\bvartheta}^{1:M}_{J + 1} = (\tilde{\vartheta}_{J+1}^1, \ldots, \tilde{\vartheta}_{J + 1}^M)$ are then updated as follows:
$$\tilde{\vartheta}^{m}_{J+1} = \tilde{\vartheta}^{m}_{J}\prod_{i=1}^n \frac{f_{\bP_{\bx_i}^{J+1}}\left(\by_i^m | \bw_{1:J+1}^m,\bpsi_{1:J+1}^m, \btheta_{1:J+1}^m\right)}{ f_{\bP_{\bx_i}^{J}}\left(\by_i^m | \bw_{1:J}^m,\bpsi_{1:J}^m, \btheta_{1:J}^m\right)}.	$$
When the effective sample size (ESS) 
is lower than a threshold, indicating poor mixing, the particle values are resampled according to such weights \citep{DelMoral_etal_2006}. Here, we resort to systematic resampling \citep{Kitagawa_1996} and perform a rejuvenating step \citep{Gilks_Berzuini_2001}, where the particles are replaced with new values sampled through $m^*$ iterations of the adaptive MCMC with $J_0 = J + 1$. The SMC provides weighted samples from the sequence of truncated posteriors $\bP^n_{J}$, converging to the infinite posterior $\bP^n$.  
To decide when a sufficiently accurate approximation has been obtained, we follow \cite{Griffin16} and stop at the  truncation level $J^*$, such that the discrepancy $D(\bP_J^n,\bP_{J+1}^n)= |\text{ESS}_{J}-\text{ESS}_{J+1}|$ is less than a specified $\delta>0$, for a fixed number $I$ of consecutive increments, $J=J^*-I+1,\ldots, J^*.$ 
We use the suggested values of $\delta=0.01M$, $I=4$, and $m^* = 3$. As an alternative to the ESS, we also consider a discrepancy based on the conditional effective sample size (CESS), which was proposed by \cite{Zhou16},
in the context of model comparison via SMC.

\paragraph{Simulation study.} To assess the performance of the proposed procedure, we applied our model to a simulated dataset with a known structure, mimicking the most relevant features of our motivating data. Specifically, we considered $q^*=3$ covariates; the first, $x_1$, is continuous and observed at a discrete scale (resembling the age at interview in our case study); the others, $x_2^*$ and $x_3^*$, are categorical with three and two levels, respectively. We generate two positive integer-valued responses, $Z_1$ and $Z_2$, and one binary response, $Z_3$, related in different extents to the covariates. $Z_1$ is a discretized noisy observation of a nonlinear function of $x_1$. Similarly, $Z_2$ is a discretized noisy observation of a nonlinear function of $x_1$ and the realized $z_1$. In both cases, the response curves are the same for $x_2^*=2,3$ and differ for other categorical combinations, while the errors are not normal but right skewed, additionally depending on $x_1$ and $x_3^*$ for the second response. Censoring is defined before discretization, when the responses are greater than the first covariate. Finally,  $Z_3$ was simulated from a linear probit model depending only on $x_1$. Complete details of the data-generating distributions are provided in the SM, together with the specification of the prior parameters. 
	
We performed a robustness analysis on the simulated dataset, comparing several initializations, namely by setting $J_0 = 2,3,5,10,15,20,30$. We found that initializing the algorithm with a conservative number of components for moderate sample sizes provides a good compromise between computational time, mixing, and accuracy. For large sample sizes however, we suggest to initialize with a generous number of components because of the computational burden implied by resampling. Specifically, for modest sample sizes, we can save the parametric mixture likelihoods, unnormalized weights, and normalizing constants for each observation and for every particle, with a memory complexity of $\mathcal{O}(nJM)$. For large sample sizes, this becomes too costly and  these terms need to be recomputed at each block update of the MCMC rejuvenation step of the SMC. 
Thus, it is convenient to minimize the need of resampling for large sample sizes, by not being too conservative in setting $J_0$ for the problem at hand.
We also considered two different discrepancy measures defining the stopping rule of the SMC, i.e. ESS and CESS, and results confirmed robustness to such choice.

Finally, focusing on the model flexibility and its ability to recover the correct structure present in the data, we performed additional simulations, exploring longer chains, increased sample size, and omitted censoring. The results, reported in the SM, are satisfactory for all scenarios, indicating (as expected) improvements when a longer chain is run or the censoring is removed.

Overall, our model was able to recover the underlying structure present in the data. 
The true conditional behavior was well recovered in areas where data is available. However, as can be expected, the model struggled when predicting at values far from the observed data. We highlight that interpretation of the conditional dependence structure in the latent scale as well as the latent covariance matrices of the mixture components and its relation to the dependence structure on the observed scale is an open and interesting direction of research, which would expand the work of \cite{garcia2007} in the parametric setting. We also observed improvements in comparison with a parametric version of our model.

\section{Application: life patterns of Colombian women}\label{sec:Application}

We employ our model to study the joint distribution of the ages at \emph{Sexual Debut} ($Z_1$), \emph{Union} ($Z_2$), and \emph{First Child} ($Z_3$) as well as \emph{Work Status} ($Z_4$) at the moment of the interview, conditional on the considered covariates. These are \emph{Age} at interview ($X_1$), \emph{Region} ($X_2^*$) and \emph{Area} ($X_3^*$) of residence, having (P) or not  (${{\bar{\mbox{P}}}}$) been disciplined using \emph{Physical Punishment} ($X_4^*$) during childhood, and having (B) or not  (${{\bar{\mbox{B}}}}$) been exposed to \emph{Parental Domestic Violence}  ($X_5^*$), referring to whether the respondent witnessed her father beating her mother.

To specify our model, we define the link functions: 
\begin{align*}
z_\ell &= h_\ell( \by,\bx)=c_\ell( \by,\bx) \lfloor\exp (y_\ell)\rfloor, \quad \text{for } \ell=1,2,\\
z_3 &= h_3( \by,\bx)=c_3( \by,\bx) \lfloor\exp (y_1) +\exp (y_3)\rfloor,\\
z_4 &= h_4(y_4, \bx)= \ind_{[0,\infty)} (y_4),
\end{align*}
with $c_\ell( \by,\bx)= \ind_{(0,x_1+1)}(\exp(y_\ell))$, for $\ell=1,2$, and $c_3( \by,\bx)= \ind_{(0,x_1+1)}(\exp (y_1) +\exp (y_3))$.  
In this case,  $\exp(y_1)$ and $\exp(y_2)$ can be interpreted as the latent continuous ages at sexual debut  and union, respectively.  
The constraint that age at first child must be greater than  age at sexual debut is strictly enforced through the transformation, and we can interpret $\exp(y_3)$ as the latent continuous time between sexual debut and first child and $\exp(y_1)+ \exp(y_3)$ as the latent continuous age at first child. The bounds required in the adaptive MCMC are obtained from inverting $z_\ell = h_\ell( \by,\bx)$; concretely, 
\begin{align*}
(l_\ell, u_\ell)& =\begin{cases}(\log(x_1+1),\infty) & \text{for censored } z_\ell=0 \\ (\log(z_\ell),\log(z_\ell+1)) & \text{for uncensored } z_\ell \neq 0
\end{cases}, \text{ when } \ell=1,2,\\
(l_3, u_3)& =\begin{cases}( \log(\max[0,x_{i,1}+1-\exp(y_{i,1})]) ,\infty) & \text{for } z_3=0 \\ (\log(\max[0,z_{i,\ell}-\exp(y_{i,1})]),\log(z_{i,\ell}-\exp(y_{i,1})+1) ) & \text{for } z_3 \neq 0
\end{cases},\\
(l_4, u_4) &= \begin{cases} (-\infty,0)& \text{for } z_4=0 \\ (0,\infty)& \text{for } z_4=1 \end{cases}.
\end{align*}

The prior parameters are specified as follows. For the linear coefficients and covariance matrix of each component, they are set empirically based on a multivariate linear regression fit. Specifically, we set $y_{i,\ell}= (l_{i,\ell}+u_{i,\ell})/2$. For $\ell=3$, when the lower bound is $-\infty$, i.e. age at sexual debut is equal to age at first child, we set $y_{i,3}= u_{i,3}-1$. For censored observations, we sample $y_{i,\ell}$ from a truncated normal distribution with mean and covariance computed from the uncensored observations. 
For the binary response,  $y_{4,i}=-1$ for $z_{4,i}=0$ and $y_{4,i}=1$ for $z_{4,i}=1$.  A multivariate linear regression fit for this auxiliary response gives estimates  $\widehat{\bbeta}$ of the linear coefficients and $\widehat{\bSigma}$ of the covariance matrix. We then define
\begin{align*}
\mathbb{E}[\bbeta_j]=\bbeta_0= \widehat{\bbeta} \quad \text{and} \quad \mathbb{E}[\bSigma_j]= \frac{1}{\nu-b-1} \bSigma_0=\widehat{\bSigma}. 
\end{align*}
Together, $\bU$ and $\bSigma_j$ reflect the variability of $\bbeta_j$ across components, and we set the matrix $\bU $ such that $\min(\diag(\widehat{\bSigma}))\,\bU=20 (\bX^\transpose \bX)^{-1}$. 
The factor $20$ for this g-prior was selected to ensure reasonable ages (i.e. mostly lower than 100) in prior simulations and to avoid very extreme age values that can result when the constant is too big (making the prior too vague). Indeed, we explored more uninformative and vague prior choices but found that this could lead to quite large and unreasonable imputed ages for censored data. We further set $\nu=b + 3$. Other specified hyperparameters include $\mu_{0,1}=\bar x_{1}$; $u_1=1/2$; $\alpha_1=2$; $\gamma_1=u_1(\text{range}(x_{1:n,1})/4)^2$; $\bvarrho_k=(1,1)$ for $k=2,\ldots,q$; and the parameters of the stick-breaking prior are $\zeta_{j,1}=1$ and $\zeta_{j,2}=1$.

We initialize the MCMC algorithm with a number of components, $J_0=35$, large enough to avoid a small ESS and subsequent resampling. The MCMC  is run for 20,000 iterations after discarding the first 30,000 as burn-in, and one in every 10 iterations is saved to produce 2,000 particles. For the SMC, we choose the ESS-based stopping rule, due to the robustness observed for the simulated data sets.

\subsection{Posterior predictive checks} 

\begin{figure}[!t]
	\begin{center}
		\includegraphics[width=0.5\textwidth]{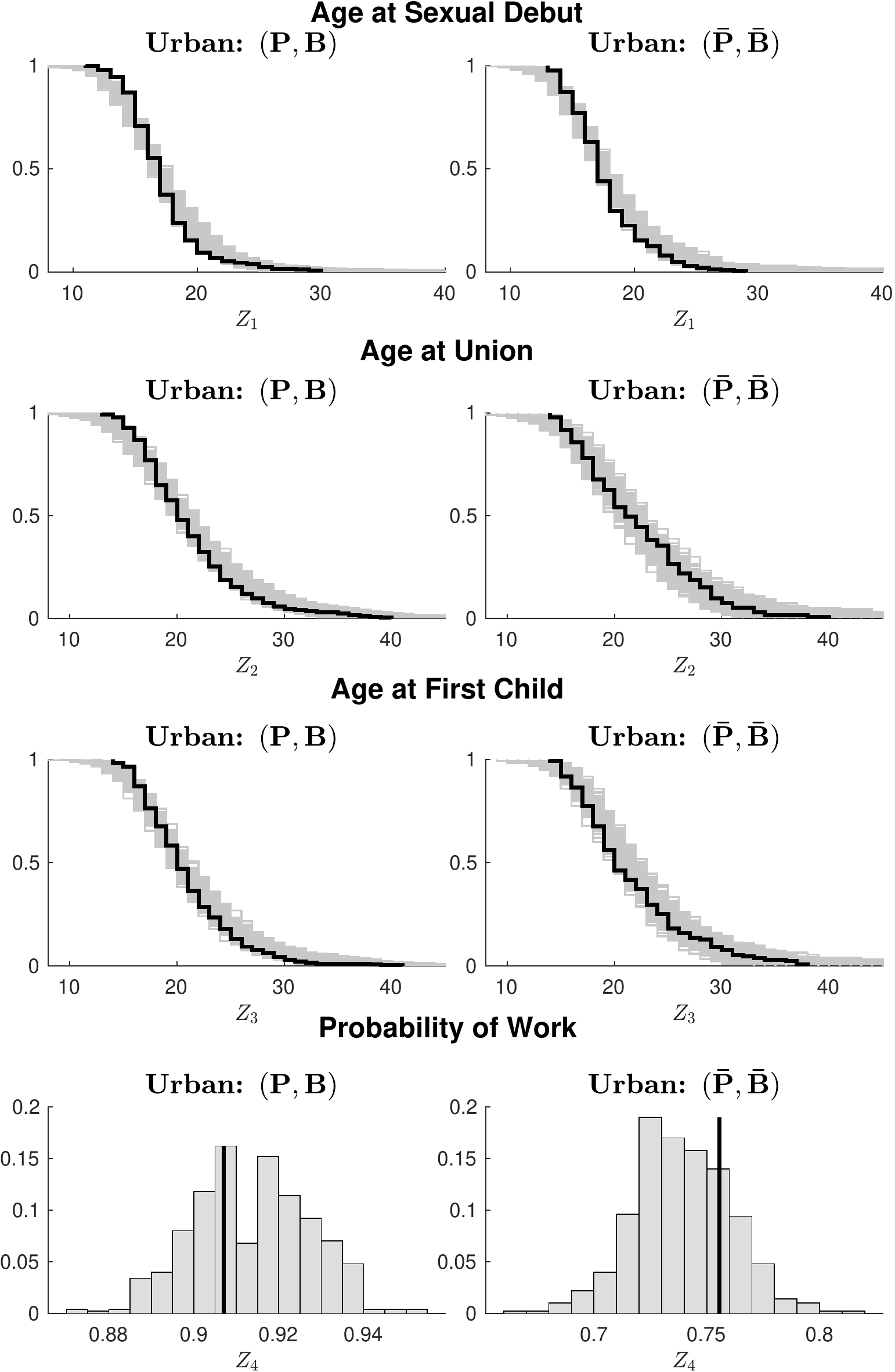}
		\caption[]{Predictive checks for the region Bogota, for women who grew up in  violent (${\mathbf{P}}, {\mathbf{B}}$) and non-violent (${\mathbf{\bar{P}}}, {\mathbf{\bar{B}}}$) families. Rows (1-3):  comparison between Kaplan-Meier curves of the discretized ages at events for the observed (black lines) and replicated (grey lines) data. Last row: comparison between proportions of working women in the observed (black vertical line) and replicated (grey histograms) data.} 
		\label{fig:PredChecks_Bogota}
	\end{center}
\end{figure}

\begin{figure}[!t]
	\begin{center}
		\includegraphics[width=0.95\textwidth]{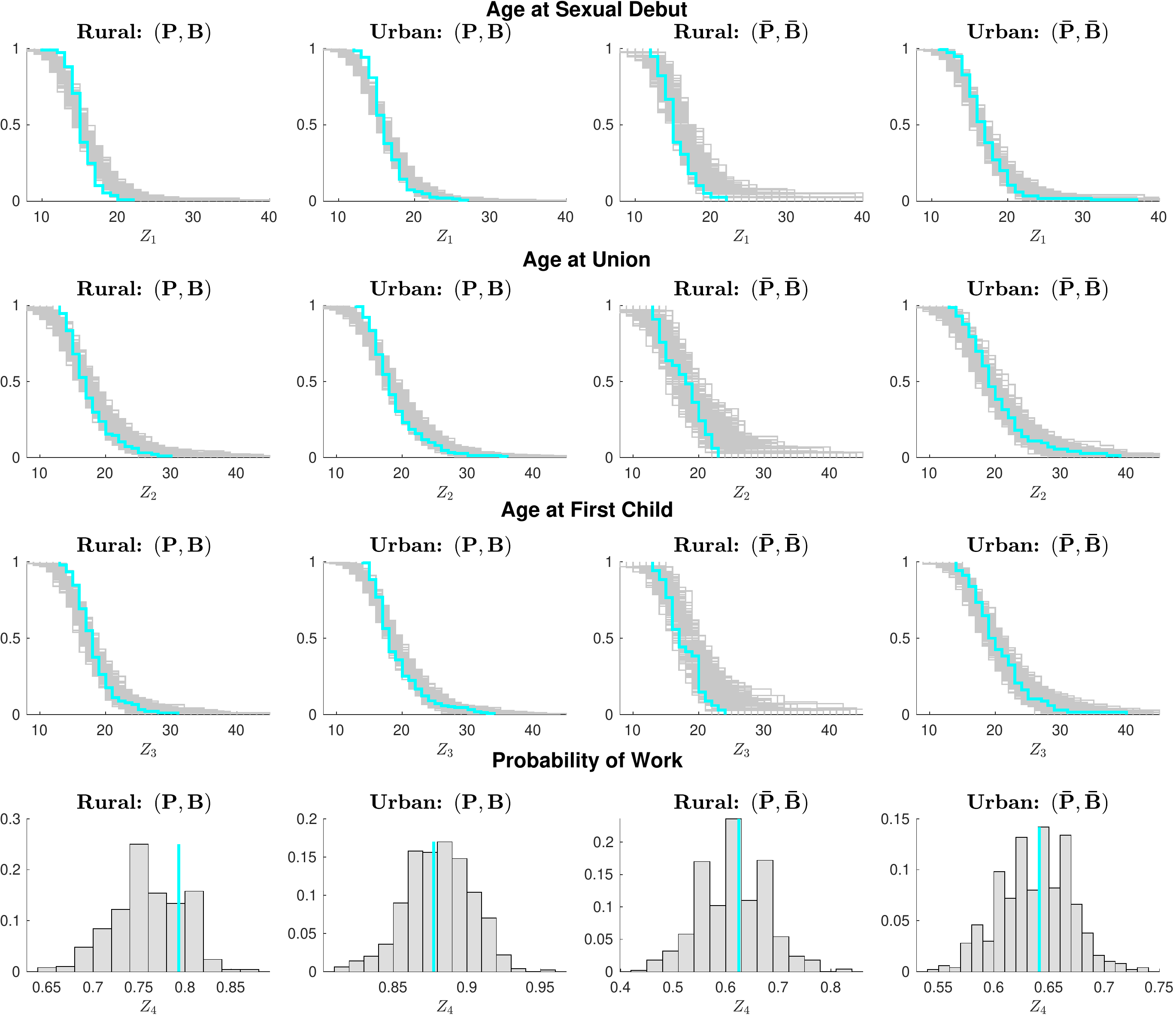}
		\caption[]{Predictive checks for the region Territorios Nacionales, in Rural or Urban areas, for women who grew up in  violent (${\mathbf{P}}, {\mathbf{B}}$) and non-violent (${\mathbf{\bar{P}}}, {\mathbf{\bar{B}}}$) families. Rows (1-3): comparison between Kaplan-Meier estimators of the discretized ages at events for the observed (light blue lines) and replicated (grey lines) data. Last row: comparison between proportions of working women in the observed (light blue vertical line) and replicated (grey histograms) data.} 
		\label{fig:PredChecks_TerrNac}
	\end{center}
\end{figure}

The assessment of the goodness of fit of the proposed model is of crucial importance when the data are of complex nature, such as the data analysed in this work. Therefore, before presenting the main results, we report posterior predictive checks in order to validate our model.  Using the weighted particles produced by the algorithm in Section \ref{sec:AdaptiveTrunc}, we first replicate the data and use the replications to check the predictive power of our model.
Figures \ref{fig:PredChecks_Bogota} and \ref{fig:PredChecks_TerrNac} report such predictive checks for the regions Bogota and Territorios Nacionales. In particular, for the discretized ages at events, the figures compare the Kaplan-Meier curves for the data and each replicated data set. For the \textit{Work} response, the histogram of the proportion of working women across the replicated datasets is displayed, along with the proportion in the observed data. In general, the estimates observed in the original data match the ones computed for the replicates, within all the considered categories, indicating good fit of the model to the data.

In addition, we compute the posterior predictive p-values \citep{Gelman_1996}:
$$
p(\bz) = \bbP(T(\bz^{rep}) \geq T(\bz) | \bz) = \int \bbP(T(\bz^{rep}) \geq T(\bz) | \bz, \bm \xi) \pi(\bm \xi | \bz)d\bm \xi,
$$
where $T(\bz^{rep})$ and $T(\bz)$ represent the selected \textit{discrepancy} computed respectively for replicated and observed data, and $\bm \xi = (\bw_{1:J^*},  \bm{\psi}_{1:J^*}, \bm{\theta}_{1:J^*} )$ indicates the model parameters with posterior $\pi(\bm \xi | \bz)$. The integral above can be computed based on the weighted particles, by  simulating a replicated dataset for each particle. In our application, the discrepancy measures used, for $\ell = 1,2,3$, are:
\begin{align*}
&T^{\text{cens}}(\bc_{\ell}) = \frac{1}{n}\sum_{i = 1}^n |1 - c_{i,\ell} - \bbP (\tilde{Z}_{\ell}\geq (x_{i,1}+1) | \bx_i,\bm{\xi}^m)|, \\
&T^{\text{noncens}}(\bz_{\ell}) = \frac{1}{|A|}\sum_{i \in A}|z_{i,\ell} - \text{median}(\tilde{Z}_{\ell}; \bm{\xi}^m)|, \quad A = \{i \in \{1, \dots, n\} | c_{i,\ell} = 1\}, \\
&T(\bz_4) = \frac{1}{n}\sum_{i = 1}^n|z_{i,4} - \bbP (Z_4 = 1| \bx_i,\bm{\xi}^m)|.
\end{align*}
Note that for the censored data, we only observe that the event is not experienced by the given age, and thus the discrepancy $T^{\text{cens}}$ is defined as for a binary variable. On the other hand, for non-censored events, the discrepancy $T^{\text{noncens}}$ can be based on the estimated age at event. The posterior predictive p-values for the full sample and for subsets defined based on selected combinations of categorical covariates are reported in Table \ref{tab:Bayesian_pvalues}. Results are encouraging and support the goodness of fit observed in Figures \ref{fig:PredChecks_Bogota} and \ref{fig:PredChecks_TerrNac}, but highlight possible issues in interpreting censoring results.

\begin{table}[!ht]
	\centering
	{\renewcommand{\arraystretch}{1.1}
		\resizebox{0.99\textwidth}{!}{
			\begin{tabular}{C{0.5cm}C{1.5cm}|C{1.5cm}|C{1.5cm}C{1.5cm}|C{1.5cm}C{1.5cm}C{1.5cm}C{1.5cm}}
			& & Full & \multicolumn{2}{c|}{Bogota} & \multicolumn{4}{c}{Terr. Nac.} \\
			& & Sample & Urban: (${\mathbf{P}}, {\mathbf{B}}$) & Urban: (${\mathbf{\bar{P}}}, {\mathbf{\bar{B}}}$) & Rural: (${\mathbf{P}}, {\mathbf{B}}$) & Urban: (${\mathbf{P}}, {\mathbf{B}}$) & Rural: (${\mathbf{\bar{P}}}, {\mathbf{\bar{B}}}$) & Urban: (${\mathbf{\bar{P}}}, {\mathbf{\bar{B}}}$) \\\hline
			\multirow{2}{*}{$Z_1$} & $c_i = 0$ & 0.148 & 0.032 & 0.885 & 0.792 & 0.989 & 0.697 & 0.826 \\
								   & $c_i = 1$ & 1 &  1 & 1 & 1 & 1 & 0.878 & 0.994 \\\hline
			\multirow{2}{*}{$Z_2$} & $c_i = 0$ & 0.056 & 0.227 & 0.843 & 0.9615 & 0.353 & 0.9945 & 0.725 \\
								   & $c_i = 1$ & 1 & 0.991 & 0.658 & 1 & 1 & 0.966 & 0.931 \\\hline
			\multirow{2}{*}{$Z_3$} & $c_i = 0$ & 0.315 & 0.591 & 0.124 & 0.671 & 0.765 & 0.998 & 0.269 \\
			 					   & $c_i = 1$ & 0.939 & 0.986 & 0.611 & 0.982 & 0.3325 & 0.483 & 0.824 \\\hline
							 $Z_4$ & & 0.604 & 0.363 & 0.935 & 0.872 & 0.547 & 0.775 & 0.604 \\\hline
			\end{tabular}
		}
	}
	\caption[]{Posterior predictive p-values for selected discrepancies computed for the full sample and subsets corresponding to violent (${\mathbf{P}}, {\mathbf{B}}$) and non-violent (${\mathbf{\bar{P}}}, {\mathbf{\bar{B}}}$) families. We report the results obtained for the regions Bogota and Territorios Nacionales.}
	\label{tab:Bayesian_pvalues}
\end{table}

\subsection{Results}
\begin{figure}[!t]
    \begin{center}
        \includegraphics[width=0.95\textwidth]{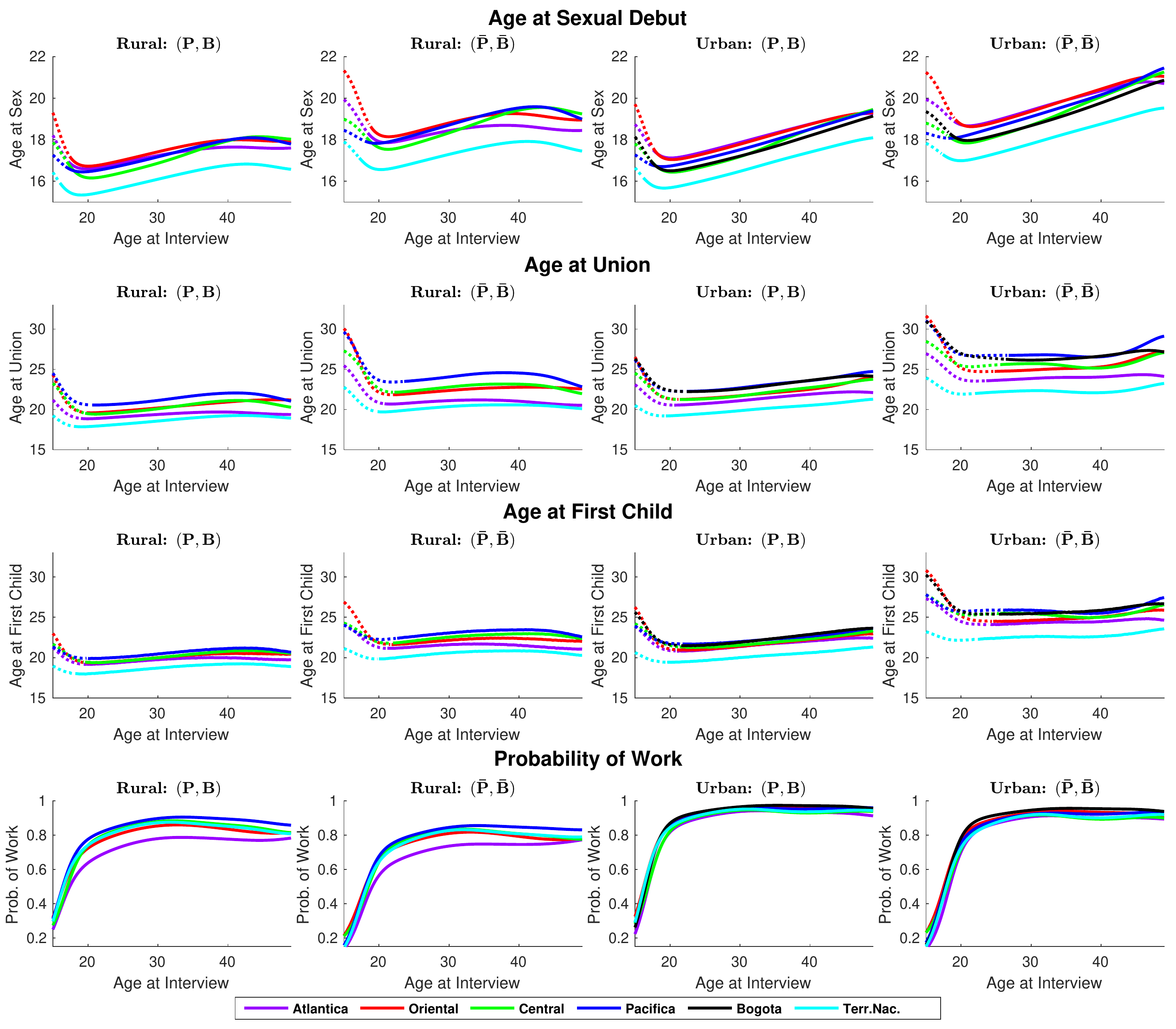}
        \caption[]{Predictive medians of the ages at sexual debut, union and first child, and posterior probability of working, as functions of \textit{Age}, for women who grew up in  violent (${\mathbf{P}}, {\mathbf{B}}$) and non-violent (${\mathbf{\bar{P}}}, {\mathbf{\bar{B}}}$) families. Dotted lines indicate when the median exceeds \textit{Age}.} 
        \label{medians_work}
    \end{center}
\end{figure}

\begin{figure}[!t]
    \begin{center}
        \includegraphics[width=0.95\textwidth]{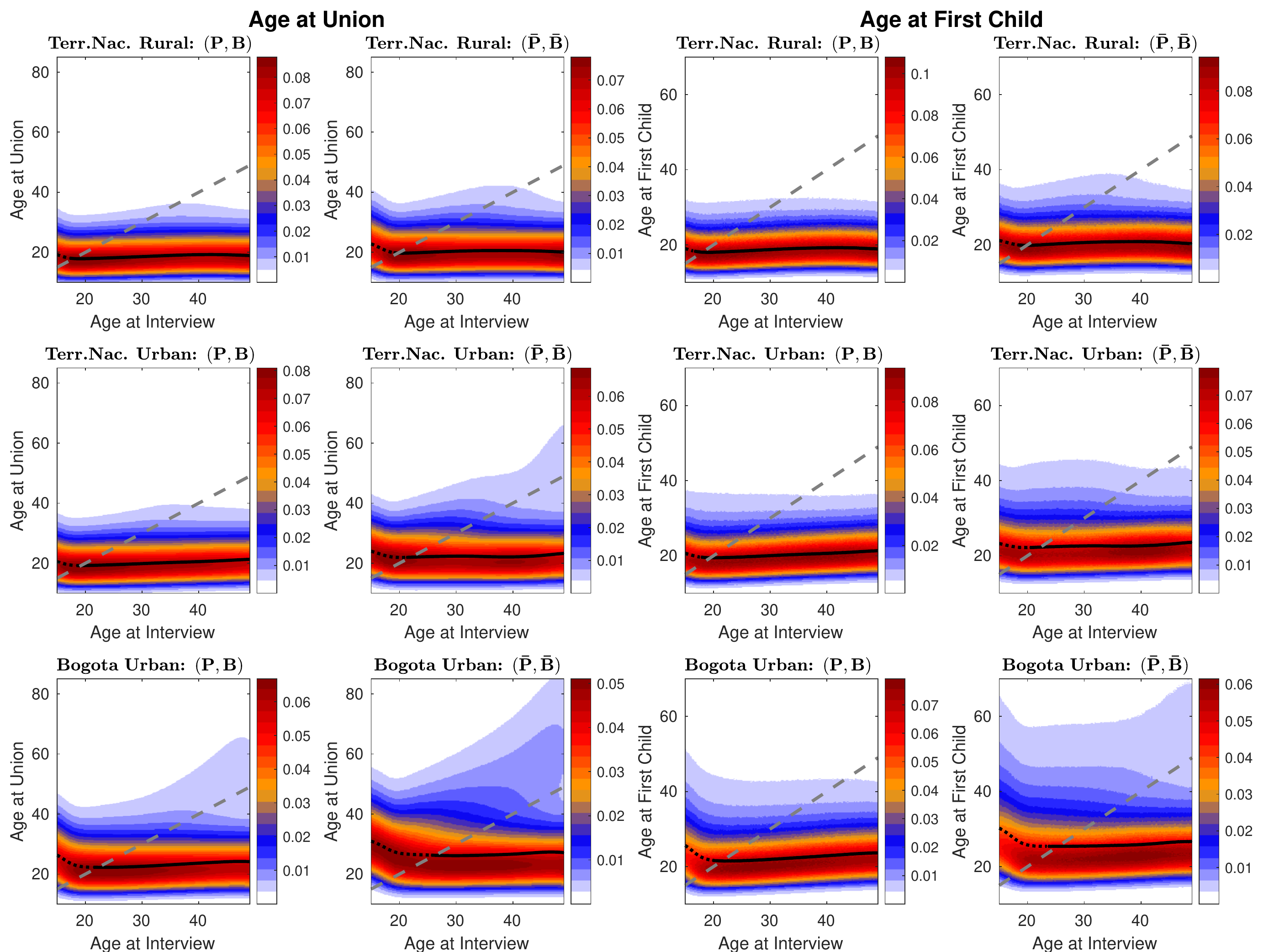}
        \caption[]{Predictive densities of the ages at union and first child as functions of \textit{Age} for women who grew up in violent (${\mathbf{P}}, {\mathbf{B}}$) and non-violent (${\mathbf{\bar{P}}}, {\mathbf{\bar{B}}}$) families. Results are reported for urban and rural areas of the least developed region (Territorios nacionales) and for the capital (Bogota). The region above the dashed line indicates when age at event exceeds \textit{Age}. The black line is the posterior median function.}
        \label{pred_densities}
    \end{center}
\end{figure}

\begin{figure}[t!]
    \begin{center}
        \includegraphics[width=0.95\textwidth]{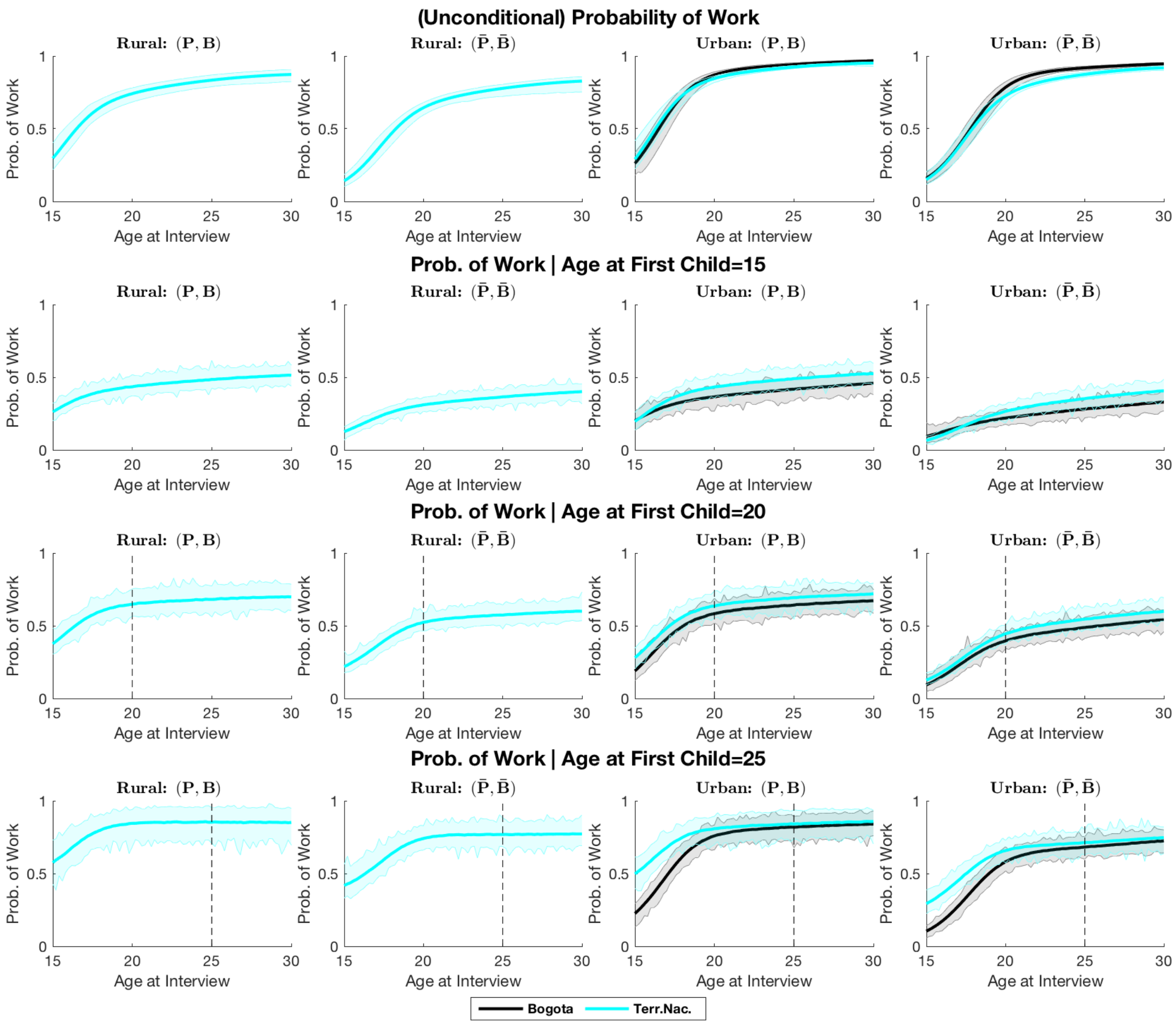}
        \caption[]{Predictive (unconditional) probability of working as function of $Age$ with 95\% pointwise credible intervals (top row) for women who grew up in violent (${\mathbf{P}}, {\mathbf{B}}$) and non-violent (${\mathbf{\bar{P}}}, {\mathbf{\bar{B}}}$) families. Subsequent rows depict the probability conditional on different ages at first child. Results are reported for urban and rural areas of the least developed region (Territorios Nacionales) and for the capital (Bogota). Left of the dashed line indicates when the age at first child is higher than \emph{Age}.}
        \label{workgivenchild_CI}
    \end{center}
\end{figure}

In the SM, we describe various posterior and predictive quantities that can be computed from the weighted particles to describe the relationship between the observed responses and the covariates. 
For the sake of conciseness, here we display only a selection of predictive quantities, which offer some insights about the situation of Colombian women. Specifically, we compare women who were raised in violent family environments (P,B) with those who were not (${{\bar{\mbox{P}}}}$,${{\bar{\mbox{B}}}}$). 
Figure~\ref{medians_work} displays the predictive medians of the (undiscretized)  ages at events  and the posterior probability of working as functions of $Age$.
More detailed information arises from the analysis of the predictive densities, some of which are reported in Figure~\ref{pred_densities}. Notice that due to the clear asymmetry in the densities, the predictive median allows a better representation of the center, as opposed to the mean. We highlight that to improve visualization we focus on predictive quantities (medians, probability of success, densities, etc.), but the Bayesian approach permits to also study uncertainty and report credible intervals for these quantities (e.g. Figure \ref{workgivenchild_CI}).

The data presents heavy censoring for younger cohorts (summarized in 
Table \ref{tbl:censoring}). 
This information is included by imputing, at each iteration of the algorithm, ages at events which must be higher than \textit{Age}. Indeed, above the dashed lines of Figure~\ref{pred_densities}, the density estimates are based on these imputed ages and borrowing of information at other covariate levels. Therefore, while we can reliably estimate the mass above the dashed line given \textit{Age}, caution should be used when interpreting the shape of the right tail in this region, as this is not identifiable from the observed data. Moreover, when such mass exceeds 0.5, the predictive median is 
affected by the imputed values and is therefore less reliable. This corresponds to median values of age at event which are higher than \emph{Age}, represented as dotted lines in the figures. 
Further, censored data also arise from women who will never experience an event. This is the prevailing cause of censoring for the older cohorts, contributing to higher medians and heavier right tails. 
Accommodating censored cases is clearly useful; however, results arising from heavily censored data should be interpreted with caution. 

Starting with Figure~\ref{medians_work}, observe that the shapes of the median curves 
change across combinations of the 
categorical covariates, which justifies the employment of a flexible model that does not impose a single functional form.
A clear difference is evident between urban and rural areas, the latter presenting lower ages at events, controlling for other covariates. This is expected since rural areas are generally characterized by lower levels of education and wealth indicators, both identified in the literature as factors related to anticipation of sexual activity and family formation.
Comparing cohorts, we observe that younger women  tend to anticipate sexual debut, a phenomenon largely recognized as a consequence of the better knowledge and the more diffuse use of contraceptive methods. 
Instead, the curves for the ages at union and at first child appear flatter, particularly for urban women with non-violent family environments and are even increasing for women from violent families. 
At first, this may seem counter-intuitive, because one would expect the younger generations to postpone family formation, particularly in urban areas, due to an expected prolonged education. However, an incorrect use of contraceptive methods, particularly among very young or less educated women, and the violent conditions linked to the armed conflict may result in early pregnancies \citep{Ali_etal, FlorezNunez_2001, Lancet}. 
Indeed, an increase in teenage childbearing in Colombia has been observed since 1990, mainly among women from disadvantaged backgrounds \citep{Batyra2016,FlorezSoto_2007, FlorezSoto_2013}. 

\begin{figure}[t!]
    \begin{center}
        \includegraphics[width=0.95\textwidth]{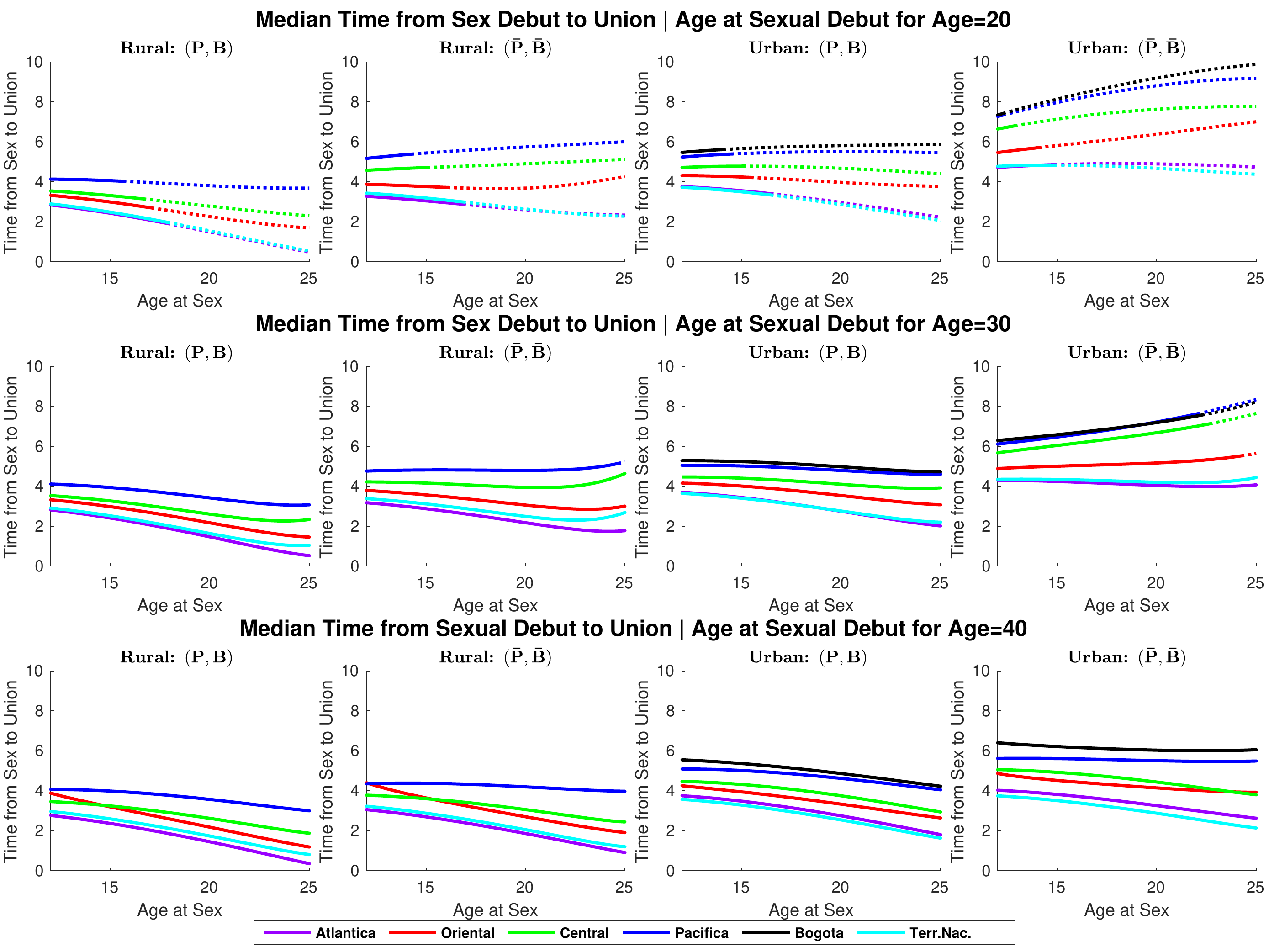}
        \caption[]{Predictive medians of the time from sexual debut to union conditional on age at sexual debut, as a function of the latter, for women  with $\textit{Age}=20, 30, 40$, who grew up in violent (${\mathbf{P}}, {\mathbf{B}}$) and non-violent families (${\mathbf{\bar{P}}}, {\mathbf{\bar{B}}}$).  
            Dotted lines indicate ages at event higher than the \textit{Age}.}
        \label{medians2_conditional}
    \end{center}
\end{figure}

Focusing on the predictive densities for the least and the most developed regions, Territorios Nacionales and the capital city Bogota (Figure~\ref{pred_densities}), further justifies the use of a density regression model. In fact, the observed flat median curves correspond to rather different distributional behaviors of ages at union and child, across covariate values. Moving from the least to the most developed context (top to bottom in the figure) entails an increase of the median curves, dispersion, and probabilities of not having experienced the events by a given \emph{Age}. An increased dispersion, with pronounced right-skewness, is more evident for older cohorts in urban environments. 
This is in line with the greater heterogeneity in urban contexts as well as with the wider range of opportunities offered, for example in terms of education. Such heterogeneity becomes more pronounced among the older cohorts who have had time to profit from such opportunities. The flexibility gained in urban contexts is offset in violent environments, thus resulting in more concentrated distributions. 
While our definition of a violent environment is not formal and refers only to the adoption of physical punishment methods and exposure to parental violence, 
the results signal the detrimental effect of family violence on  Colombian women life patterns, and provides quantitative support for psychosocial studies \citep{martinez2017sexual}.

The joint modeling approach permits us to study also the conditional relation between responses. For example, Figure~\ref{medians2_conditional} shows the conditional predictive medians of the time from sexual debut to union given the age at sexual debut  for women with  $\textit{Age}=20, 30, 40$ (dotted lines indicate predicted ages at event higher than  \textit{Age}; the corresponding conditional densities are reported in the SM).  Interesting differences can be observed across regions, likely related to their socio-demographic characteristics 
\citep{Profamilia_2011_DHS}. We observe that women in  Atlantica and Territorios Nacionales (and to a lesser extent Oriental) compared with Pacifica and Bogota tend to experience sexual debut and union closer in time, suggesting that sexual debut is possibly delayed until union. Such tendency is more pronounced, compared to the other regions, for rural women raised in violent families. 
Similar results are observed for the time from sexual debut to child (details in the SM).

\begin{figure}[t!]
    \begin{center}
        \includegraphics[width=0.95\textwidth]{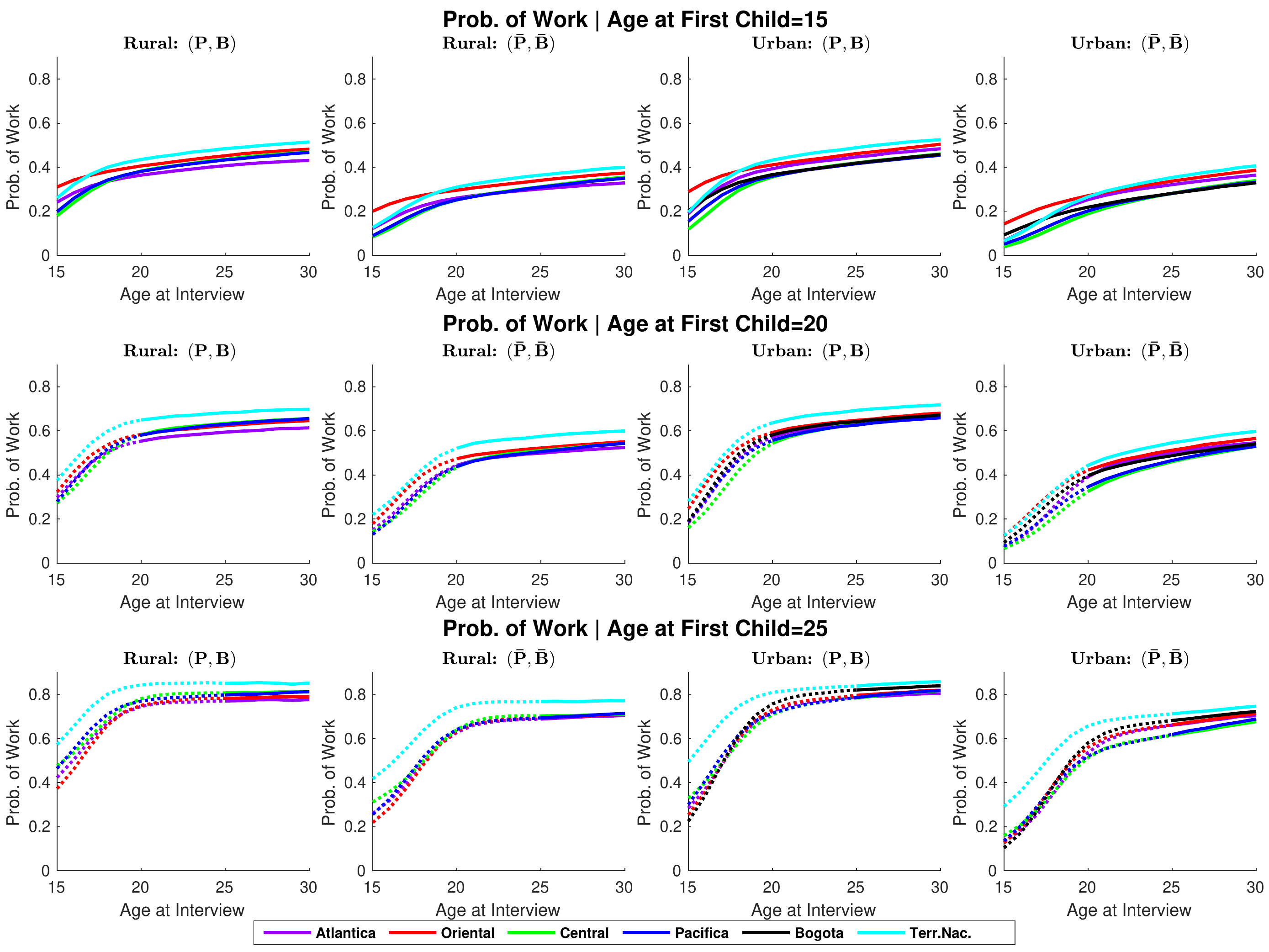}
        \caption[]{Predictive probability of working as function of $Age$ conditional on different ages at first child, for women who grew up in violent (${\mathbf{P}}, {\mathbf{B}}$) and non-violent (${\mathbf{\bar{P}}}, {\mathbf{\bar{B}}}$) families. Dotted lines indicate ages at event higher than \emph{Age}.}
        \label{workgivenchild}
    \end{center}
\end{figure}

Finally, the probability of working (Figure~\ref{medians_work}, bottom row) is, as expected, higher in urban areas.  Moreover, women who grew up in violent environments show a higher propensity to work, more pronounced among younger women. These same women, as previously observed, show a tendency to anticipate events. 
A possible explanation is that young women who leave the parental house to escape violence may start cohabitation and decide to drop out of school, entering the labor market to contribute to family income.  
This apparently contradicts studies \citep[see e.g.][]{Gimenez_etal_2015} pointing to the difficulties of young women, especially those with children, to participate in the labor market. However, this paradox is solved when analyzing the estimated predictive probabilities of working as functions of $Age$, conditional on having the first child at ages 15, 20 and 25 (Figure~\ref{workgivenchild}, top to bottom). Indeed, the probability of working at each \emph{Age} increases with the age at first child. In particular, we observe a much lower probability of working for young mothers, that persists even when considering their labor market participation later in life. This suggests a scaring effect of teenage motherhood. This effect is further supported by Figure \ref{workgivenchild_CI}, which depicts the 95\% pointwise credible intervals for the probability of working as functions of $Age$, comparing the unconditional probability (top row) and the probability conditional on different ages at first child (subsequent rows). Consistent with previous results, we focus on the least developed region (Territorios Nacionales) and on the capital (Bogota).

\section{Concluding remarks}\label{sec:Conclusions}
In this work, we proposed a novel Bayesian nonparametric model for density regression, allowing for mixed responses with censored, constrained, and binary traits, that can flexibly change with combinations of the categorical and numerical covariates. We developed a general algorithm for posterior inference, that effectively scales to large datasets by adaptively determining the necessary truncation level to approximate the infinite-dimensional posterior. We customized the model and algorithm to a specific case study, but they can be applied in other contexts through minor modifications, by appropriate definition of the link functions. Note that our model accommodates for non-informative (or random) censoring. Interesting extensions concern other types of censoring.
From a technical point of view, our results highlight the advantage of a flexible model, accounting for a different shape, location, and dispersion of the response distribution across the covariate levels, as well as for censoring. Additionally, a variety of classic graphic tools and quantities of interest, such as survival curves and hazard functions, can be derived. Importantly, the joint analysis of the responses allows for a rich variety of conditional analyses, which can be conducted focusing on different aspects, a very useful feature when studying complex phenomena.

For our case study, the findings suggest interesting considerations regarding life patterns of Colombian women. 
In the first instance, we found a confirmation of the differences  between rural and urban areas, which evidence the need of interventions towards a more balanced development of the country. Furthermore, our results signal that the regions with a higher risk of early transition to adulthood are those with the worse development and wellness indicators, thus corroborating studies on the risks related to disadvantageous conditions. 
One of the most interesting results is the rather clear evidence of the impact of family violence on women's choices and behaviors. An anticipation of the considered events is observed for women who were physically punished during childhood and witnessed parental domestic violence, two factors we used as proxies for a violent family environment. The relation between child abuse and neglect and the child's future family choices has been discussed in the literature. Nonetheless, to our knowledge, this is the first attempt to study the possible relation between  parental family violence and the events marking the transition to adulthood. Our findings confirm  that a violent family environment can be regarded as a key risk factor that may nullify the positive influence of developed areas. 

Overall, our case study may contribute to the planning of targeted interventions. Even if recent governments have shown an increased attention to the conditions of women and children,
a  formal statistical approach to systematically identify and quantify critical situations is crucial to support such a process.
For example, teenage pregnancy is recognized as a priority issue in Colombia by the Government \citep{Gimenez_etal_2015, Lancet}, 
due to its hindering personal development and agency \citep{Azevedo_2012}; our results confirm its scaring effect and quantify the risk of teenage pregnancy, identifying  some of the most vulnerable groups. We conclude with the hope that the present work may stimulate further reflection, research and survey on the topic, and possibly lead to additional investigations exploiting the availability of DHS surveys on other developing countries and the flexibility and wide applicability of our model.


\bibliographystyle{ba}
\bibliography{ref_Combined}

\begin{acknowledgement}
The work reported in this paper was funded by the University of Warwick Academic Returners Fellowship and the University of Oslo. Raffaella Piccarreta acknowledges MIUR-bando PRIN 2017 for the support in her contribution to the final article. \\
The authors are grateful to the Editor, the Associate Editor  and the referees for their thoughtful and helpful comments on a previous version of the manuscript.

See 
the Supplementary Material for further details. 
The code is publicly available, along with the simulated data to reproduce results:\\
 \url{https://github.com/sarawade/BNPDensityRegression_AdaptiveTruncation} \\ 
\end{acknowledgement}

\end{document}


\begin{frontmatter}
    \title{Supplementary Material for ``Colombian Women's Life Patterns: A Multivariate Density Regression Approach''}
    \runtitle{SM for Multivariate Density Regression}
\begin{aug}
	\author{\fnms{Sara} \snm{Wade}\thanksref{addr1}\ead[label=e1]{sara.wade@ed.ac.uk}},
	\author{\fnms{Raffaella} \snm{Piccarreta}\thanksref{addr2}\ead[label=e2]{raffaella.piccarreta@unibocconi.it}},
	\author{\fnms{Andrea} \snm{Cremaschi}\thanksref{addr4}\ead[label=e3]{andrea.cremaschi@yale-nus.edu.sg}},
	\and
	\author{\fnms{Isadora} \snm{Antoniano-Villalobos}\thanksref{addr5, addr3}
		\ead[label=e4]{isadora.antoniano@unive.it}} 
	
	\runauthor{S. Wade et al.}
	
	\address[addr1]{School of Mathematics, University of Edinburgh, Edinburgh 
		, UK 
		\printead{e1} 
	}
	
	\address[addr2]{Department of Decision Sciences, Dondena Research Centre, Bocconi Institute for Data Science and Analytics (BIDSA), Bocconi University, 
		Milan, Italy
		\printead{e2}
	}
	
	\address[addr4]{Yale-NUS College, Singapore
		\printead{e3}
	}
	
	\address[addr5]{Department of Environmental Sciences, Informatics and Statistics, Ca' Foscari University of Venice and BIDSA, Italy
		\printead{e4}
	}

	
	
\end{aug}
	   
\end{frontmatter}
 
\appendix
\addappheadtotoc
\setcounter{table}{0}
\setcounter{figure}{0}
\renewcommand{\thetable}{\Alph{section}.\arabic{table}}
\renewcommand{\thefigure}{\Alph{section}.\arabic{figure}}

In this Supplementary Material, we describe how to compute  predictive quantities of interest from the MCMC output in Section \ref{app:Predictions}, and  in Sections \ref{app:simulated} and \ref{app:application}, we report additional results for the simulated data example and the application to the Colombian women dataset described in the paper.


\section{Predictions}\label{app:Predictions}

The weighted posterior samples obtained with the adaptive truncation algorithm can be used to produce various posterior and predictive quantities of interest. Here, we describe how to compute the predictive densities, medians, \textit{probability of censoring} for  the (undiscretized) ages at event and probability of success for binary variables. Full details and implementation for other quantities are provided in the  accompanying software and documentation: 
\url{https://github.com/sarawade/BNPDensityRegression_AdaptiveTruncation}.

Focusing on the application in Section 6, for $\ell=1,2,3$, we denote by $\tilde{Z}_\ell$ the (undiscretized) age at sexual debut, the (undiscretized) age at union, and the time from sexual debut to first child, respectively. These are linked to our model by the relation $\tilde{Z}_\ell= \exp(Y_\ell)$, and the corresponding ages are obtained through discretization. The (undiscretized) age at first child is  denoted as $ \breve{Z}_{3}= \tilde{Z}_{1}+\tilde{Z}_{3}$. For \emph{Work Status}, we have $Z_{4}=\ind_{(0,\infty)}(Y_{4})$. 
In the following, let $J$ denote the final truncation level, with corresponding weighted particles $(\bw_{1:J}^m,\btheta_{1:J}^m,\bpsi_{1:J}^m, \by_{1:n}^m)$ and unnormalized particle weights $\tilde{\vartheta}^m$, for $m=1,\ldots, M$ (without loss of generality, we drop the subscript $J$). We indicate with $\vartheta^m$, for $m=1,\ldots, M$, the normalized particle weights.
%

We begin with marginal predictive quantities of interest.
First, the predictive probability of success for a binary response given $\bx_*$ (e.g.  for $\ell =4$, shown in Figure 5, bottom row) is: 
\begin{align*}
\bbP (Z_{*,\ell}=1| \bx,\bz,\bx_*)
=\bbP(Y_{*,\ell}>0| \bx,\bz,\bx_*) \approx   \sum_{m=1}^M \vartheta^m \sum_{j=1}^J w_j^m(\bx_*) \Phi\left(\frac{\bm \bx_*\bbeta_{j,(\cdot,\ell)}^m}{\sqrt{\bSigma_{j,(\ell,\ell)}^{m}}}\right).
\end{align*}
For $\ell=1,2,3$, 
the marginal predictive density of 
$\tilde{Z}_{*,\ell}$ given $\bx_*$, shown in Figure 6 for some values of $\bx_*$, is given by:
\begin{align}
f(\tilde{z}_{*,\ell}| \bx,\bz, \bx_*)&\approx \sum_{m=1}^M \vartheta^m \sum_{j=1}^J w_j^m(\bx_*) f(\tilde{z}_{*,\ell}| \btheta_j^m, \bx_*) \nonumber \\
&= \sum_{m=1}^M \vartheta^m \sum_{j=1}^J w_j^m(\bx_*) \lN(\tilde{z}_{*,\ell}|\bx_*\bbeta_{j,(\cdot,\ell)}^m,\bSigma_{j,(\ell,\ell)}^{m}),  \label{eq:margpreddens}
\end{align}
for $\tilde{z}_{*,\ell}>0$, where $\bbeta_{j,(\cdot,\ell)}^m$ denotes the $\ell$-th column of $\bbeta$ in component $j$ and particle $m$;  $\bSigma_{j,(\ell,\ell)}^{m}$ denotes element $(\ell,\ell)$ of the matrix $\bSigma$ in component $j$ and particle $m$; and $\lN(\cdot|\mu,\sigma^2)$ denotes the log-normal density with parameters $\mu$ and $\sigma^2$. 
Due to the skewness of the predictive densities in our application (Section 6), we focus on the predictive median over the predictive mean to better represent the central tendencies and summarize the predictive densities. 
The marginal predictive median (Figure 5) can be computed numerically by evaluating the marginal predictive density \eqref{eq:margpreddens} on a sufficiently dense grid of $\tilde{z}_{*,\ell}$ values.  


For $\ell=1,2$ corresponding to age at sexual debut and union, an interesting quantity is the predictive probability that the indexed event has not yet occurred for a new individual with $x_{*,1}$ years of age (Figure \ref{censorprob}), computed as:
\begin{align}
\bbP (\tilde{Z}_{*,\ell}\ge (x_{*,1}+1)&|\bx,\bz, \bx_*)
= \bbP (Y_{*,\ell}> \log(x_{*,1}+1)|\bx,\bz, \bx_*)\nonumber\\
&\approx  \sum_{m=1}^M \vartheta^m \sum_{j=1}^J w_j^m(\bx_*) \left(1- \Phi\left(\frac{ \log(x_{*,1}+1)- \bx_*\bbeta_{j,(\cdot,\ell)}^m}{ \sqrt{\bSigma_{j,(\ell,\ell)}^{m}}} \right)\right).\label{eq:censoring}
\end{align}
This can be interpreted as the \textit{predictive probability of censoring} of the event for a new individual and corresponds to the mass above the dashed line of Figure 6, given $x_{*,1}$.

Our model also recovers the joint relationship between responses, which allows inference on conditional properties. Specifically, when $\ell$ indexes a binary response and $\ell'$ indexes an age at event response, the conditional predictive probability of success given $\tilde{z}_{*,\ell'}$ and $\bx_*$ (Figure 8) is:
\begin{small}\begin{align*}
\bbP (Z_{*,\ell}=1| \tilde{z}_{*,\ell'},\bx,\bz,\bx_*)
\approx   \sum_{m=1}^M \vartheta^m \sum_{j=1}^J w_j^m(\bx_*) \Phi\left(\frac{\mu_{j,\ell|\ell'}^m}{\sqrt{\sigma_{j,\ell|\ell'}^{2\, m}}}\right) \frac{\lN(\tilde{z}_{*,\ell'}|\bx_*\bbeta_{j,(\cdot,\ell')}^m,\bSigma_{j,(\ell',\ell')}^{m}) }{f(\tilde{z}_{*,\ell'}|\bx,\bz, \bx_*)},
\end{align*}\end{small}
where  $$\mu_{j,\ell|\ell'}^m= \bx_*\bbeta_{j,(\cdot,\ell)}^m+\bSigma_{j,(\ell,\ell')}^{m}(\bSigma_{j,(\ell',\ell')}^{m})^{-1}(\log(\tilde{z}_{*,\ell'})-\bx_*\bbeta_{j,(\cdot,\ell')}^m),$$ 
$$\sigma_{j,\ell|\ell'}^{2\, m}=\bSigma_{j,(\ell,\ell)}^{m}-(\bSigma_{j,(\ell,\ell')}^{m})^2(\bSigma_{j,(\ell',\ell')}^{m})^{-1},$$ and the density in the denominator is the marginal predictive of equation \eqref{eq:margpreddens}. 
For $\ell\neq\ell'$ both indexing ages at event, the conditional predictive density of $\tilde{Z}_{*,\ell}$ given $\tilde{z}_{*,\ell'}$ and $\bx_*$ takes the form: 
\begin{align}
&f(\tilde{z}_{*,\ell}|\tilde{z}_{*,\ell'}, \bx,\bz, \bx_*)=  \nonumber \\&\quad \sum_{m=1}^M \vartheta^m \sum_{j=1}^J w_j^m(\bx_*)
\lN(\tilde{z}_{*,\ell}| \mu_{j,\ell|\ell'}^m,\sigma_{j,\ell|\ell'}^{2\, m})
\frac{\lN(\tilde{z}_{*,\ell'}|\bx_*\bbeta_{j,(\cdot,\ell')}^m,\bSigma_{j,(\ell',\ell')}^{m}) }{f(\tilde{z}_{*,\ell'}|\bx,\bz, \bx_*)}. \label{eq:condpreddens}
\end{align}
Figure \ref{app_density_conditional_sextounion} shows the conditional predictive density of $\tilde{Z}_{*,2}-\tilde{z}_{*,1}$ given $\tilde{z}_{*,1}$ and $\bx_*$, which can be easily computed from \eqref{eq:condpreddens}. The corresponding  predictive medians (Figure 7) can be obtained numerically from evaluations of this density on an adequate, dense grid of 
values. The conditional density plot for $\tilde{Z}_{*,3}$ given $\tilde{z}_{*,1}$ (Figure \ref{app_density_conditional_sextochild}) and the corresponding median (Figure \ref{app_medians1_conditional}) can be obtained directly from equation \eqref{eq:condpreddens}.

Finally, we note that for the (undiscretized) age at first child, $ \breve{Z}_{*,3}= \tilde{Z}_{*,1}+\tilde{Z}_{*,3}$, and more generally constrained responses, the corresponding marginal and conditional predictive quantities may require integration over $\tilde{Z}_{*,1}$. 
For example, the conditional predictive density of $\breve{Z}_{*,3}$ given $\tilde{z}_{*,1}$ is simply the conditional predictive density of equation \eqref{eq:condpreddens}, evaluated at $\breve{z}_{*,3}-\tilde{z}_{*,1}$.
%
While the marginal predictive density of $\breve{Z}_{*,3}$ given $\bx_*$ (Figure 6) is obtained as:
\begin{align}
&f(\breve{z}_{*,3} | \bx,\bz, \bx_*)= \int f(\tilde{z}_{*,3}|\tilde{z}_{*,1}, \bx,\bz, \bx_*) f(\tilde{z}_{*,1}| \bx,\bz, \bx_*) d\tilde{z}_{*,1} \nonumber \\
&\quad \approx \sum_{m=1}^M \vartheta^m \sum_{j=1}^J w_j^m(\bx_*)
\int_{-\infty}^{\breve{z}_{*,3}} \lN(\tilde{z}_{*,3}| \mu_{j,3|1}^m,\sigma_{j,3|1}^{2\, m})
\lN(\tilde{z}_{*,1}|\bx_*\bbeta_{j,(\cdot,1)}^m,\bSigma_{j,(1,1)}^{m}) d\tilde{z}_{*,1}, \label{eq:margpreddenschild}
\end{align}
where $\tilde{z}_{*,3}=\breve{z}_{*,3}-\tilde{z}_{*,1}$.
We evaluate the integral stochastically, via a Monte Carlo approximation, and compute the marginal predictive median of the undiscretized age at first child (Figure 5) numerically from the marginal predictive density in \eqref{eq:margpreddenschild}.  
Also, the predictive probability that the woman has not yet had a child at $x_{*,1}$ years of age (Figure \ref{censorprob}) takes the form: 
\begin{small}\begin{align*}
&\bbP (\breve{Z}_{*,3}>x_{*,1}|\bx,\bz, \bx_*)=\bbP (\tilde{Z}_{*,3}+\tilde{Z}_{*,1}\geq x_{*,1}+1|\bx,\bz, \bx_*)\\
&\quad \approx \sum_{m=1}^M \vartheta^m \sum_{j=1}^J w_j^m(\bx_*)
\int   \left( 1-\Phi \left(\frac{l(x_{*,1}+1)-\mu_{j,3|1}^m }{ \sqrt{\sigma_{j,3|1}^{2\,m}}}  \right) \right)
\lN(\tilde{z}_{*,1}|\bx_*\bbeta_{j,(\cdot,1)}^m,\bSigma_{j,(1,1)}^{m}) d\tilde{z}_{*,1},
\end{align*}\end{small}
where $l(z)= \log(\max(0,z-\tilde{z}_{*,1}))$. 
The conditional predictive density of the (undiscretized) age at first child $\breve{Z}_{*,3}$ given the (undiscretized) age at union $\tilde{z}_{*,2}$ and $x_{*,1}$ is:
{\small \begin{align}
&f(\breve{z}_{*,3}|\tilde{z}_{*,2}, \bx,\bz, \bx_*) \approx \sum_{m=1}^M \vartheta^m \sum_{j=1}^J w_j^m(\bx_*)  f(\breve{z}_{*,3}|\tilde{z}_{*,2}, \btheta^m_j, \bx_*) \frac{\lN(\tilde{z}_{*,2}|\bx_*\bbeta_{j,(\cdot,2)}^m,\bSigma_{j,(2,2)}^{m})}{f(\tilde{z}_{*,2}| \bx,\bz, \bx_*)}.
\label{eq:denscond_z3_z2}
\end{align}}
Notice that this expression differs from equation \eqref{eq:condpreddens} in that
\begin{small}\begin{align*}
f(\breve{z}_{*,3}|\tilde{z}_{*,2}, \btheta^m_j, \bx_*) = \int_{-\infty}^{\breve{z}_{*,3}}  \lN(\breve{z}_{*,3}-\tilde{z}_{*,1}|\mu_{j,3|(1,2)}^m,\sigma_{j,3|(1,2)}^{2\, m}) \lN(\tilde{z}_{*,1}|\mu_{j,1|2}^m,\sigma_{j,1|2}^{2\, m}) d\tilde{z}_{*,1},
\end{align*}\end{small}
where
\begin{align}
\begin{split}
	\mu_{j,3|(1,2)}^m & = \bx_*\bbeta_{j,(\cdot,3)}^m+\bSigma_{j,(3,1:2)}^{m}\bSigma_{j,(1:2,1:2)}^{-1 \,m}(\log(\tilde{z}_{*,1:2})-\bx_*\bbeta_{j,(\cdot,1:2)}^m),\\ 
\sigma_{j,3|(1,2)}^{2\, m} & = \bSigma_{j,(3,3)}^{m}-\bSigma_{j,(3,1:2)}^{m}\bSigma_{j,(1:2,1:2)}^{-1 \,m}\bSigma_{j,(1:2,3)}^{m}.
\end{split}\label{eq:uglymusigma}
\end{align}
Figure \ref{app_density_uniontochild} shows the conditional predictive density of $\breve{Z}_{*,3}-\tilde{z}_{*,2}$ given $\tilde{z}_{*,2}$ and $\bx_*$, which can be easily computed from \eqref{eq:denscond_z3_z2}, with  the
 corresponding  predictive medians in Figure \ref {app_medians2_conditional}. 
Lastly, 
the conditional predictive probability of success for a binary response, e.g. $\ell=4$ in our application, given $\breve{z}_{*,3}$ and $\bx_*$ is:
\begin{align*}
\bbP (Z_{*,4}=1| \breve{z}_{*,3},\bx,\bz,\bx_*)
\approx   \sum_{m=1}^M \vartheta^m \sum_{j=1}^J w_j^m(\bx_*)  \bbP(Y_{*,4}>0|\breve{z}_{*,3}, \btheta^m_j,\bx_*) \frac{f(\breve{z}_{*,3}| \btheta^m_j,\bx_* ) }{f(\breve{z}_{*,3}|\bx,\bz, \bx_*)},
\end{align*}
where
\begin{small}\begin{align*}
&\bbP(Y_{*,4}>0|\breve{z}_{*,3}, \btheta^m_j,\bx_*) f(\breve{z}_{*,3}| \btheta^m_j,\bx_* ) \\
&\quad=\int_{-\infty}^{\log(\breve{z}_{*,3})}  \Phi\left(\frac{\mu_{j,4|(1,3)}^m}{\sqrt{\sigma_{j,4|(1,3)}^{2\, m}}}\right) \lN(\breve{z}_{*,3}-\tilde{z}_{*,1}| \mu_{j,3|1}^m,\sigma_{j,3|1}^{2\, m})
\lN(\tilde{z}_{*,1}|\bx_*\bbeta_{j,(\cdot,1)}^m,\bSigma_{j,(1,1)}^{m}) d\tilde{z}_{*,1},
\end{align*}\end{small}
where $\mu_{j,4|(1,3)}$ and $\sigma_{j,4|(1,3)}^{2}$ are calculated analogously to expression \eqref{eq:uglymusigma}.

\section{Simulation study}\label{app:simulated}
We assess the performance of the proposed procedure on a simulated data set including $q^*=3$ covariates and $d=3$ responses. The first covariate mimics \emph{Age} and, as such, is assumed to be registered at a discrete level: $x_{1}=\lfloor\tilde{x}_1 \rfloor$, where $\tilde{x}_{1} \sim \U(15,30)$. The remaining covariates are categorical: $x_2^*$ has three levels with probabilities $(0.5, 0.3, 0.2)$ while $x_3^*$ has two levels with probabilities $(0.4, 0.6)$.

\begin{figure}[!t]
	\includegraphics[width=0.99\textwidth]{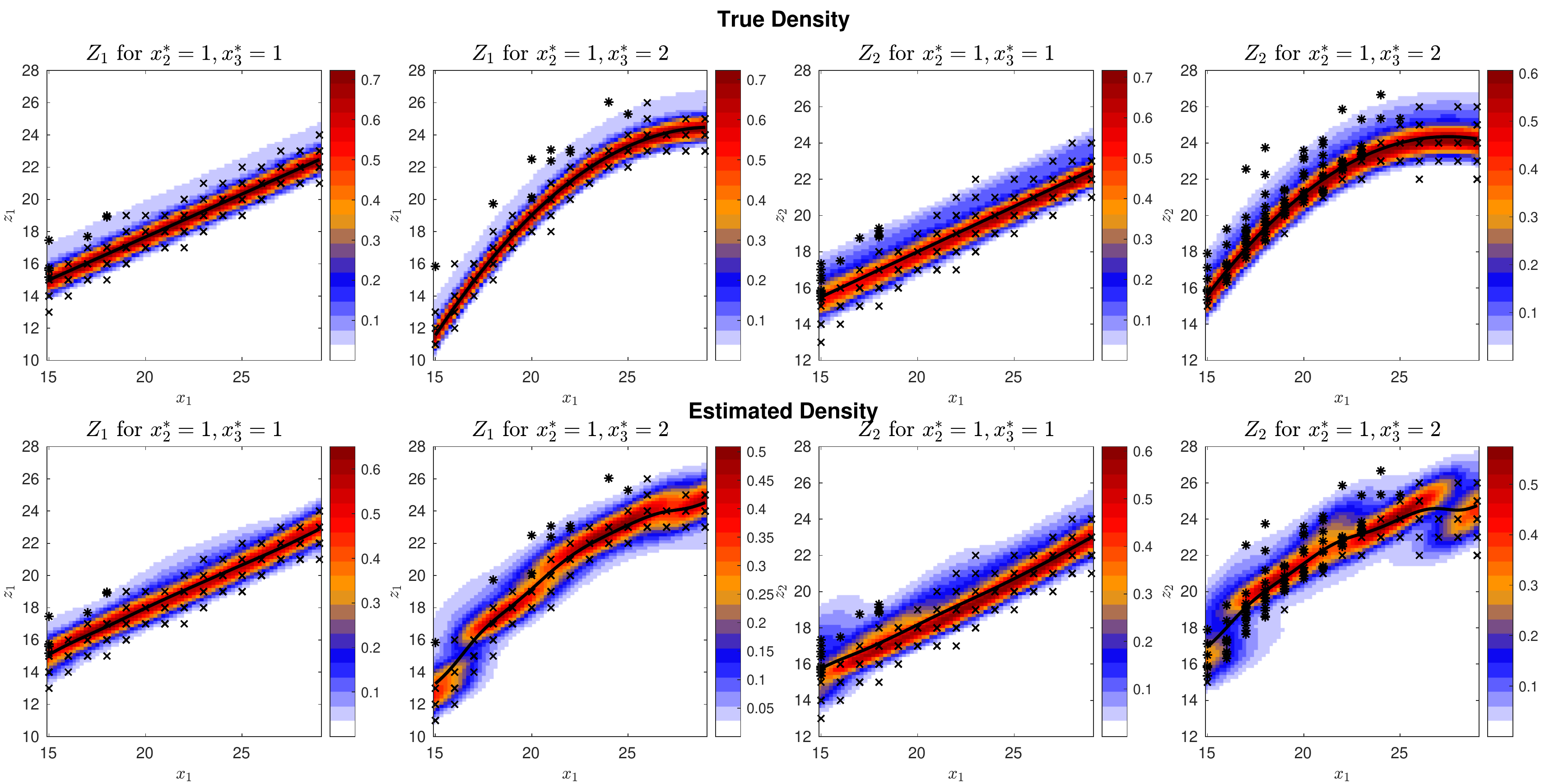} 
	\caption[]{Simulated data set. True data-generating density (top row) and estimated predictive density (bottom row) of the (undiscretized) $Z_1$ and $Z_2$ as functions of $x_1$ for two combinations of the categorical covariates. The estimated/true mean function is depicted with a black solid line; crosses and stars mark respectively observed and censored points.}
	\label{fig:zdensity}
\end{figure}

\begin{figure}[!t]
	\centering
	\includegraphics[width=0.7\textwidth]{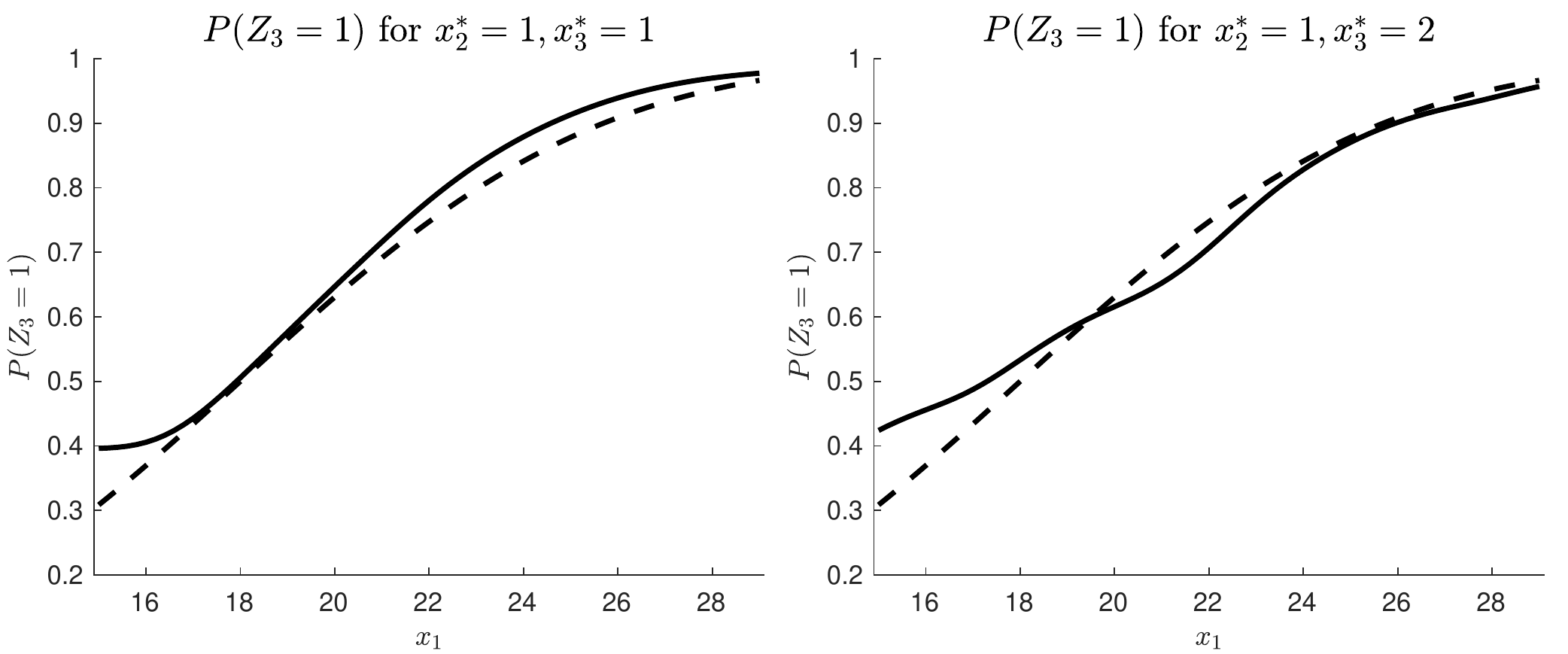}
	\caption[]{Simulated data set. True (dashed line) and predictive (solid line) probability of $Z_3=1$ as a function of $x_1$ for two combinations of the categorical covariates.}
	\label{fig:Z3Probs}
\end{figure}

We generate two positive discretized responses and one binary response.
The first response, $Z_1$, is a discretized noisy observation of a nonlinear function of $x_1$. To build $Z_1$, we first generate:
$$\tilde{Z}_{i,1}= \mu^t_1(\tilde{x}_{i,1},x^*_{i,2},x^*_{i,3}) + \epsilon_{i,1}, \mbox{ for } i = 1, \dots, n,$$
where $\epsilon_{1,1}, \dots, \epsilon_{n,1} \iidsim 0.9\N(-15/90,0.5^2)+0.1\N(1.5,0.75^2)$, and
\begin{small}\begin{equation*}
\mu^t_1(\tilde{x}_{i,1},x^*_{i,2},x^*_{i,3}) = \left\{
\begin{array}{ll}
-0.057\tilde{x}_{i,1}^2+3.08\tilde{x}_{i,1}-21.247 & \mbox{if } x^*_{i,2} \ne 1, x^*_{i,3}=2 \\
\frac{1}{3}\tilde{x}_{i,1}+10 & \mbox{if } x^*_{i,2} \ne 1, x^*_{i,3}=1  \\
0.0001\tilde{x}_{i,1}^3-0.0695\tilde{x}_{i,1}^2+3.83\tilde{x}_{i,1}-30.584 & \mbox{if } x^*_{i,2}= 1, x^*_{i,3}=2 \\
\frac{8}{15}\tilde{x}_{i,1}+7 & \mbox{if } x^*_{i,2}= 1, x^*_{i,3}=1
\end{array}
\right. .
\end{equation*}\end{small}
Similarly, $Z_2$ is a discretized noisy observation of a nonlinear function of $x_1$ and the realized $z_1$ and it is built by generating:
\begin{small}\begin{equation*}
\tilde{Z}_{i,2}= \left\{
\begin{array}{ll}
-0.056\tilde{x}_{i,1}^2+3.08\tilde{x}_{i,1}-18 +0.75\left[\tilde{z}_{i,1}-\mu^t_1(\tilde{x}_{i,1},x^*_{i,2},x^*_{i,3}) \right] + \epsilon_{i,2} & \mbox{if } x^*_{i,3}=2 \\
0.5\tilde{x}_{i,1}+8 +0.75\left[\tilde{z}_{i,1}-\mu^t_{1}(\tilde{x}_{i,1},x^*_{i,2},x^*_{i,3}) \right] + \epsilon_{i,2} & \mbox{if } x^*_{i,3}=1
\end{array}
\right.,
\end{equation*}\end{small} 
where the errors are assumed to depend also on $\tilde{x}_1$ and $x^*_{3}$:
\begin{equation*}
\epsilon_{i,2} \sim \left\{
\begin{array}{ll}
0.9\N(-\frac{1}{6},0.4^2)+0.1\N(1.5,0.75^2)  & \mbox{if } x^*_{i,3}=2 \\
0.9\N\left(-\frac{1}{6},\left(\frac{7.5}{\tilde{x}_{i,1}}\right)^2\right)+0.1\N\left(1.5,\left(\frac{7.5}{\tilde{x}_{i,1}}\right)^2\right) & \mbox{if } x^*_{i,3}=1
\end{array}
\right. .
\end{equation*}
Note that the response curves are the same for $x_2^*=2,3$ and differ for other categorical combinations, while the errors are not normal but right skewed, additionally depending on $x_1$ and $x_3^*$ for the second response. Observed responses are set to missing for censored observations, defined as individuals with $\tilde{z}_{1,i}>\tilde{x}_{1,i}$ or $\tilde{z}_{2,i}>\tilde{x}_{1,i}$.  
Since the age-related variables in our motivating application are registered at a discrete level, the observed responses were rounded down to the nearest integer, i.e. $z_1=\lfloor\tilde{z}_1 \rfloor$, $z_2=\lfloor\tilde{z}_2 \rfloor$. 
The true curves and densities are depicted in Figure~\ref{fig:zdensity} (top row) for selected combinations of the covariates.
The behavior of $Z_1$ and $Z_2$ may at a first sight seem a simple function of $x_1$ alone, however it depends also on the interactions between $x_1$ and the categorical covariates. Moreover, this consideration holds both for the regression function and for the variance. 
Combined together these aspects, which are of the same nature as those present in our motivating data, pose challenges for parametric and semi-parametric models. Indeed, the relationship between each response and the $Age$ at interview is relatively smooth, but interactions with the categorical covariates and changes in variability increase complexity. Finally, a binary response is simulated from a linear probit model depending only on $x_1$ (Figure~\ref{fig:Z3Probs}):
\begin{align*}
Z_{3,i} \sim \Bern\left(\Phi\left( \frac{\tilde{x}_{1,i}-18}{6} \right)\right).
\end{align*}

We seek to recover the conditional distribution of the response variables given the covariates using our proposed model, from a sample of size $n=700$. We define the link functions $h_\ell( \by,\bx)$ as: 
	\begin{align}
	\begin{split}
	z_\ell &= h_\ell( \by,\bx)=c_\ell( \by,\bx) \lfloor\exp (y_\ell)\rfloor, \quad \text{for  } \ell= 1,2,\\
	z_3 &= h_3(\by, \bx)= \ind_{[0,\infty)} (y_3),
	\end{split}\label{eq:linksim}
	\end{align}
	where $c_\ell( \by,\bx)= \ind_{(0,x_1+1)}(\exp(y_\ell))$. In this case, the bounds required in the adaptive MCMC are obtained from inverting $z_\ell = h_\ell( \by,\bx)$; concretely, 
	\begin{align*}
	(l_\ell, u_\ell)& =\begin{cases}(\log(x_1+1),\infty) & \text{for censored } z_\ell=0 \\ (\log(z_\ell),\log(z_\ell+1)) & \text{for uncensored } z_\ell \neq 0
	\end{cases}, \text{ when } \ell=1,2,\\
	(l_3, u_3) &= \begin{cases} (-\infty,0)& \text{for } z_3=0 \\ (0,\infty)& \text{for } z_3=1 \end{cases}.
	\end{align*}

\paragraph{Prior specification and computational details.}
Prior parameters for the linear coefficients and covariance matrix of each component are specified empirically based on multivariate linear regression fit to the data. Specifically, for $\ell = 1,2$, we set $y_{i,\ell}= (l_{i,\ell}+u_{i,\ell})/2$ and $y_{i,\ell} = \log(x_{i,1}+2)$ for uncensored and censored observations, respectively, where the bounds $l_{i,\ell}$ and $u_{i,\ell}$ are defined in Section 4. 
Additionally, we let $y_{i,3}=-1$ for $z_{i,3}=0$ and $y_{i,3}=1$ for $z_{i,3}=1$. 
A multivariate linear regression fit on these auxiliary responses gives estimates  $\widehat{\bbeta}$ of the linear coefficients and $\widehat{\bSigma}$ of the covariance matrix. We then define
\begin{align*}
\bbE[\bbeta_j]=\bbeta_0= \widehat{\bbeta} \quad \text{and} \quad \bbE[\bSigma_j]= \frac{1}{\nu-b-1} \bSigma_0=\widehat{\bSigma}. 
\end{align*}
Together, $\bU$ and $\bSigma_j$ reflect the variability of $\bbeta_j$ across components, and we set the matrix $\bU $ such that $\min(\diag(\widehat{\bSigma}))\,\bU=10 (\bX^\transpose \bX)^{-1}$. The factor of this g-prior was selected to ensure reasonable ages (i.e. mostly lower than 100) in prior simulations. We explored more uninformative and vague prior choices but found that this could lead to quite large and unreasonable imputed ages for censored data. We further set $\nu=b+3$, to ensure the existence of the first and second moments of $\bm \Sigma_j$ a-priori. Other specified hyperparameters include $\mu_{0,1}=\overline{x}_{1}$, $u_1=1/2$, $\alpha_1=2$, $\gamma_1=u_1(\text{range}(x_{1:n,1})/4)^2$, $\bvarrho_k=(1,1)$ for $k=p+1,\ldots,q$, and the parameters of the stick-breaking prior are $\zeta_{j,1}=1$ and $\zeta_{j,2}=1$. Here $\overline{x}_{1}$ and $\text{range}(x_{1:n,1})$ denote the sample mean and range of $(x_{1,1}, \dots, x_{n,1})$.

The MCMC stage of the adaptive truncation algorithm, with $J_0 = 15$ components, is run for 20,000 iterations after discarding the first 10,000 as burn-in. Every 10-th iteration is saved to produce $M $= 2,000 initial values for the particles in the SMC stage.

\begin{figure}[!h]
	\centering
	\includegraphics[width=0.9\textwidth]{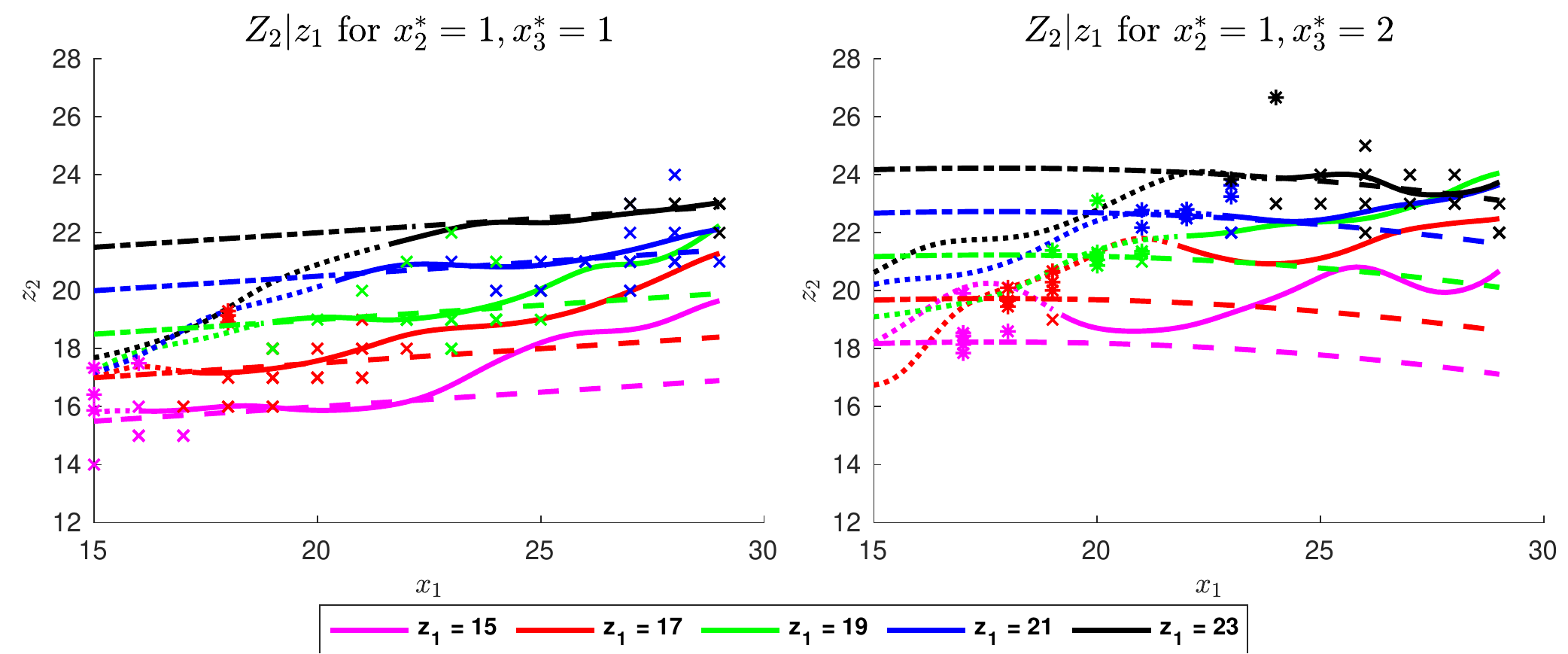}
	\caption{Simulated data set. Conditional mean of $Z_2$ given $z_1$ as a function of $x_1$ with colors representing the different values of $z_1$, for two combinations of the categorical covariates. The estimated and true conditional mean functions are depicted with solid and dashed lines respectively, and dotted lines indicate when the mean exceeds $x_1$. Crosses and stars mark respectively the observed and censored points, colored by the observed value of $z_1$. }
	\label{fig:sim_2cond1}
\end{figure}

\paragraph{Results}
In Appendix \ref{app:Predictions}, we  describe various posterior and predictive quantities that can be computed from the weighted particles to describe the relationship between the observed response $\bz$ and covariates $\bx$.
Here, we focus on the marginal predictive mean and density functions for (undiscretized) $Z_1$ and $Z_2$, as well as on the marginal predictive probability of success for $Z_3$, and compare them with the true data-generating functions in Figures~\ref{fig:zdensity} and \ref{fig:Z3Probs}, for a selected combinations of the categorical covariates. We also show the conditional mean of $Z_2$  as a function of $x_1$ given different values of $z_1$ in Figure \ref{fig:sim_2cond1}. Overall, the model is able to recover the underlying structure present in the data, despite the heavy censoring of $Z_2$ for lower levels of $x_1$, particularly when $x_3^*=2$. The true conditional structure is well recovered in areas where data (crosses) is available, i.e. in the left plot of Figure \ref{fig:sim_2cond1}, for $15 \leq x_1 \leq 18$ given $z_1 = 15$ (magenta) and for $27 \leq x_1 \leq 29$ given $z_1 = 23$ (black). However, as can be expected, the model struggles when predicting at values far from the observed data, i.e in the left plot of  Figure \ref{fig:sim_2cond1}, for $25 \leq x_1 \leq 30$ given $z_1 = 15$ (magenta) and for $15 \leq x_1 \leq 20$ given $z_1 = 23$ (black). We highlight that interpretation of the conditional dependence structure in the latent scale as well as the latent covariance matrices of the mixture components and its relation to the dependence structure on the observed scale is an open and interesting direction of research, which would expand the work of \cite{garcia2007} in the parametric setting.

\paragraph{Robustness analysis.}
We perform a robustness analysis comparing several initialization specifications, namely by setting $J_0 =2,3,5,10,15,20,30$. 
We also compare with a parametric version of the model that is similar in nature to the parametric model of \cite{Korsgaard2003},  i.e. a multivariate Gaussian regression model with the link functions $ h_\ell( \by,\bx)$ in \eqref{eq:linksim}  and a prior given by the base measure $\bP_0$. For the sake of comparison, we use the Metropolis-within-Gibbs scheme for inference. In all scenarios, the adaptive MCMC algorithm is run for 30,000 iterations, discarding the first 10,000 as burn-in, and saving only every 10th iteration for a total of $M = $ 2,000 particles to be used in the SMC step. A summary of the analysis is reported in Table \ref{tab:Errors_grid_Simul}. Besides the number of components inferred by the model ($J^*$), and the elapsed CPU time (in hours), the table reports the ESS for the log-likelihood in the MCMC stage ($\text{ESS}_{\text{MCMC}}$) computed with the \textit{mcmcse} package in \textbf{R} \citep{mcmcse}, and the $\text{ESS}_{J^*}$ of the final iteration of the SMC, that can be used to compare mixing. Results are reported for two different discrepancy measures used to define the stopping rule of SMC, namely the ESS and the CESS \citep{Zhou16}. In order to assess the fit of the model, we compute for each $Z_\ell$ the log-pseudo marginal likelihood \citep[LPML,][]{Geisser_Eddy_1979}, and the percentage absolute errors with respect to the true mean and true density at a set of new test covariates, $\bx_i^*$, for $i = 1,\ldots,n^*$, denoted by $\text{ERR}_{\text{Mean}}$ and $\text{ERR}_{\text{Dens}}$:
\begin{small}\begin{align*}
&\text{LPML}^{\ell}= \sum_{i=1}^n \log( \text{CPO}^\ell_i ) \quad \text{with} \quad  \text{CPO}^\ell_i = \left( \frac{1}{M}\sum_{m=1}^M \frac{1}{f(z_{i,\ell}|\bw^m,\bpsi^m,\btheta^m,\bx_i)} \right)^{-1}, \\
& \text{ERR}^{\ell}_{\text{Mean}} = \frac{100}{n^*}\sum\limits_{i=1}^{n^*} \frac{|\mu_\ell^t(\bx_i^*)-\widehat{\mu}_\ell(\bx_i^*)|}{|\mu_\ell^t(\bx_i^*)|}, \\
& \text{ERR}^{\ell}_{\text{Dens}} = \frac{100}{n^*}\sum\limits_{i=1}^{n^*} \frac{\int |f^t(z_{\ell}^*|\bx_i^*)-\widehat{f}(z_{\ell}^*|\bx_i^*)| dz^*_\ell}{ \int |f^t(z_{\ell}^*|\bx_i^*)|dz^*_\ell} \approx  \frac{100}{n^*}\sum\limits_{i=1}^{n^*} \sum_{g=1}^{G} |f^t(z_{g,\ell}^*|\bx_i^*)-\hat{f}(z_{g,\ell}^*|\bx_i^*)| \Delta,
\end{align*}\end{small}
where for each response $\ell=1,\ldots,d$, $\mu_\ell^t(\bx_i^*)$ and $\widehat{\mu}_\ell(\bx_i^*)$ indicate the true and estimated mean functions, and $f^t(\cdot|\bx_i^*)$ and $\widehat{f}(\cdot|\bx_i^*)$ indicate the true and estimated densities. For each response, densities are evaluated on a grid of values, $z_{1,\ell}^*,\ldots, z_{G,\ell}^*$, with grid size $\Delta$. The results show robustness with respect to the choice of the discrepancy measure.

\begin{table}[!ht]
	{\renewcommand{\arraystretch}{1.1}
		\resizebox{0.99\textwidth}{!}{
		 	\begin{tabular}{C{2cm}@{\hspace{1\tabcolsep}}C{.7cm}@{\hspace{1\tabcolsep}}C{0.7cm}@{\hspace{1\tabcolsep}}C{1cm}C{1.75cm}@{\hspace{1\tabcolsep}}C{1.5cm}@{\hspace{1\tabcolsep}}C{1.05cm}@{\hspace{1\tabcolsep}}C{1.05cm}@{\hspace{1\tabcolsep}}C{1.05cm}@{\hspace{1\tabcolsep}}C{1.05cm}@{\hspace{1\tabcolsep}}C{1.05cm}@{\hspace{1\tabcolsep}}C{1.05cm}@{\hspace{1\tabcolsep}}C{1.15cm}@{\hspace{1\tabcolsep}}C{1.05cm}@{\hspace{1\tabcolsep}}C{1.05cm}}
				 & $J_{0}$ & $J^{*}$ & CPU & $\text{ESS}_{\text{MCMC}}$ & $\text{ESS}_{J^*}$ & \multicolumn{3}{c}{LPML ($10^3$)} & \multicolumn{3}{c}{$\text{ERR}_{\text{Mean}}$} & \multicolumn{3}{c}{$\text{ERR}_{\text{Dens}}$} \\
				& & & & & & $Z_1$ & $Z_2$ & $Z_3$ & $Z_1$ & $Z_2$ & $Z_3$ & $Z_1$ & $Z_2$ & $Z_3$ \\\hline
				 Parametric & 1 & 1 & 0.66 & 767.3 &  & -1.17 & -0.82 & -0.34 & 5.82 & 3.26 & 15.39 & 163.79 & 115.32 & 13.86 \\ \hline
				$\text{ESS}_J$ & 2 & 13 & 1.90 & 495.0 & 1,125.5 & -1.01 & -0.76 & -0.34 & 3.93 & 5.38 & 6.99 & 132.55 & 119.69 & 7.23 \\ 
				& 5 & 14 & 3.93 & 519.6 & 1,966.7 & -0.91 & -0.71 & -0.34 & 2.94 & 4.89 & 7.41 & 97.80 & 104.74 & 6.59 \\ 
				& 10 & 19 & 5.29 & 537.0 & 1,918.9 & -0.94 & -0.71 & -0.34 & 2.41 & 3.12 & 8.42 & 90.47 & 98.56 & 7.99 \\ 
				& 15 & 26 & 5.77 & 543.5 & 1,266.0 & -0.93 & -0.73 & -0.35 & 2.94 & 3.75 & 9.30 & 82.99 & 109.44 & 8.32 \\ 
				& 20 & 24 & 5.43 & 567.8 & 1,990.3 & -0.86 & -0.69 & -0.34 & 2.31 & 3.56 & 7.92 & 78.89 & 95.45 & 7.88 \\ 
				& 30 & 34 & 9.04 & 550.8 & 2,000.0 & -0.85 & -0.69 & -0.34 & 2.59 & 3.67 & 8.37 & 78.06 & 97.60 & 7.96 \\ \hline
				$\text{CESS}_J$ & 2 & 14 & 3.67 & 495.0 & 1,989.8 & -1.01 & -0.76 & -0.34 & 3.89 & 5.26 & 6.82 & 129.35 & 116.34 & 7.09 \\ 
				& 5 & 14 & 3.90 &  519.6 & 1,978 & -0.91 & -0.71 & -0.34 & 2.94 & 4.89 & 7.41 & 97.80 & 104.79 & 6.59 \\ 
				& 10 & 17 & 5.18 & 537.0 & 1,905.9 & -0.92 & -0.71 & -0.34 & 2.35 & 2.92 & 8.21 & 89.79 & 98.81 & 7.96 \\ 
				& 15 & 23 & 6.13 & 543.5 & 1,974.9 & -0.93 & -0.73 & -0.35 & 3.00 & 3.75 & 9.35 & 84.40 & 109.41 & 8.38 \\ 
				& 20 & 24 & 5.51 & 567.8 & 1,994.1 & -0.86 & -0.69 & -0.34 & 2.31 & 3.56 & 7.92 & 78.89 & 95.45 & 7.88 \\ 
				& 30 & 34 & 11.17 & 550.8 & 2,000 & -0.85 & -0.69 & -0.34 & 2.59 & 3.67 & 8.37 & 78.06 & 97.60 & 7.96 \\  \hline
			\end{tabular}
				}
			}
	\caption{Simulated data set. Summaries of the performance: computational burden, mixing, goodness of fit, and predictive errors in mean and density obtained with the parametric model (first row) and 
	 the nonparametric model for different values of $J_0$. Results are reported for the adaptive truncation algorithm based on the $\text{ESS}$ and $\text{CESS}$ stopping rules.}
	\label{tab:Errors_grid_Simul}
\end{table}

\begin{figure}[!h]
	\centering
	\includegraphics[width=0.9\textwidth]{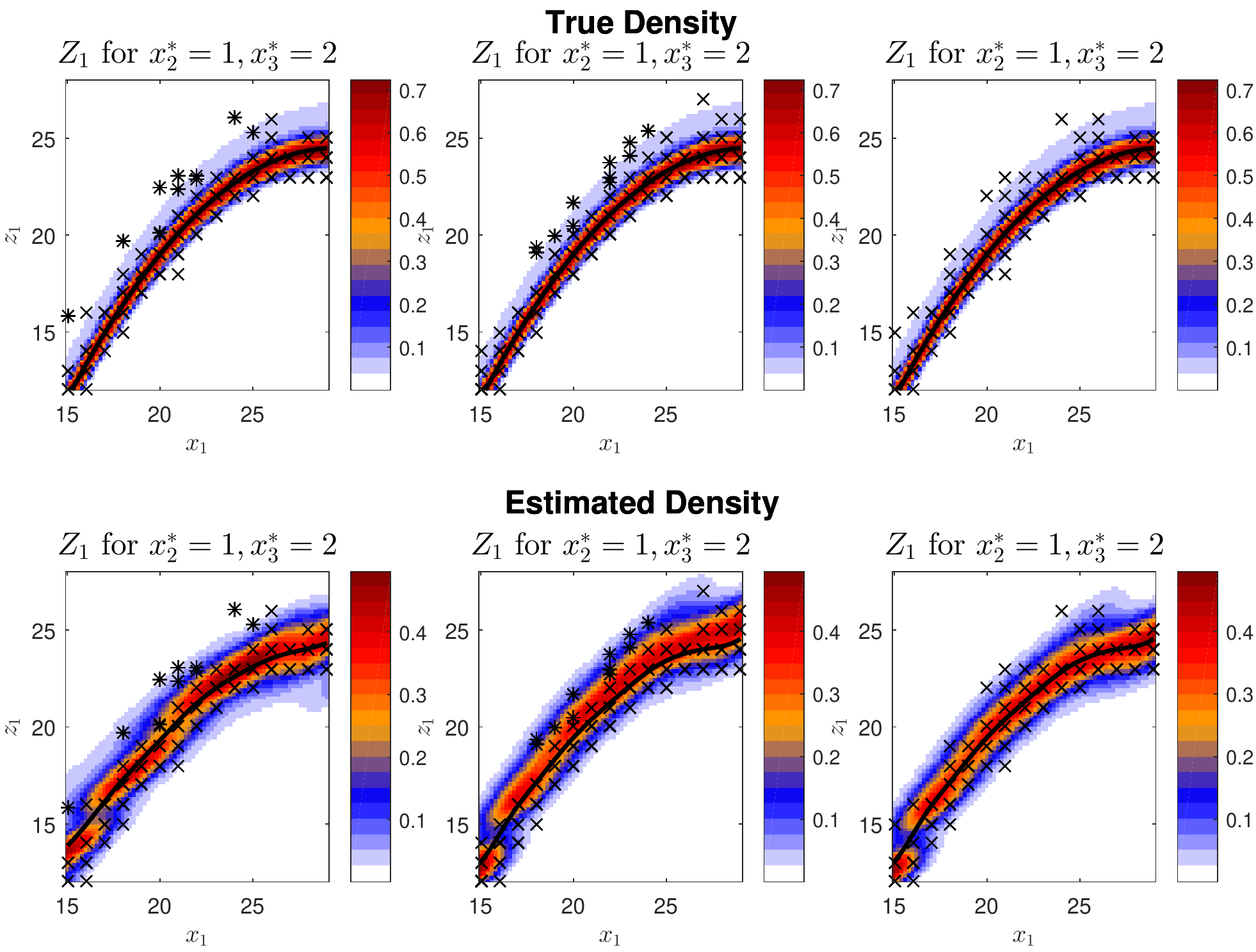}
	\caption{Simulated data set. Alternative scenarios by column: (i) longer MCMC chain; (ii) $n = 1,400$; (iii) no censoring. True data-generating density (top) and estimated predictive density (bottom) of the (undiscretized) $Z_1$ as a function of $x_1$ for one combination of the categorical covariates. The true and estimated mean functions are depicted with black solid lines; crosses and stars mark observed and censored points, respectively.}
	\label{fig:NewSimulations_DensityFig}
\end{figure}

We observe that for $J_0\geq20$ only a moderate number of components are added, suggesting that a sufficient approximation is obtained with around 20 components. Recall that the SMC is run for at least $I=4$ cycles, i.e. at least four new components are added to the initial model. Therefore, if $J_0$ is large enough, we have $J^*= J_0+I$. 
Generally, the computational time increases with $J_0$, although this is not always  the case, especially if $\text{ESS}_{J}$  becomes too low so that resampling and rejuvenation are required. Despite the increased number of parameters for large $J_0$, the mixing of the MCMC, reflected in the $\text{ESS}_{\text{MCMC}}$, does not deteriorate; however, note the improved mixing for the parametric model, which has the least number of parameters, due to the absence of the covariate-dependent weights. Focusing on the SMC, a larger $J_0$ generally results in less degeneracy of the particles, reflected in a higher $\text{ESS}_{J^*}$. Finally the LPML, measuring the goodness of fit of the model,  increases with $J_0$, while the errors in predictive mean and density both decrease. This is particularly true for $Z_1$, the most nonlinear response, while there is little improvement in the binary response $Z_3$, which is indeed simulated from a linear probit model. Similar results are obtained when substituting the ESS with the CESS in the discrepancy measure of the SMC, confirming robustness to the choice of the stopping rule. 
To conclude, initializing the algorithm with a conservative number of components provides a good compromise between computational time, mixing, and accuracy.

Focusing on the model flexibility and its ability to recover the correct structure present in the data, we explore three additional scenarios: (i) longer number of iterations (25,000 burn-in; 10 thinning; $M = 2,500$); (ii) doubled sample size ($n = 1,400$); (iii) omitted censoring. In all cases, $J_0 = 15$. The results are reported in Table \ref{tab:Errors_grid_Simul_2}, showing the same goodness-of-fit and error indicators used in Table \ref{tab:Errors_grid_Simul}.
As expected,  mixing improves for scenario (i), due to the higher number of iterations, and modest improvements in the predictive power are observed. Increasing the sample size (scenario (ii)) without adjusting the algorithm settings (e.g., length of the chain, $J_0$), does not lead to consistent improvements. Note that this scenario yields approximately doubled LPML values, due to the definition of this quantity. The removal of censoring (scenario (iii)) yields faster and more precise computations. Estimates of the predictive densities for the undiscretized variable $Z_1$ as function of $x_1$ are reported in Figure \ref{fig:NewSimulations_DensityFig}.

\begin{table}[!t]
	\centering
	{\renewcommand{\arraystretch}{1.1}
		\resizebox{0.99\textwidth}{!}{
			\begin{tabular}{C{.7cm}@{\hspace{1\tabcolsep}}C{0.7cm}@{\hspace{1\tabcolsep}}C{1cm}C{1.75cm}@{\hspace{1\tabcolsep}}C{1.5cm}@{\hspace{1\tabcolsep}}C{1.05cm}@{\hspace{1\tabcolsep}}C{1.05cm}@{\hspace{1\tabcolsep}}C{1.05cm}@{\hspace{1\tabcolsep}}C{1.05cm}@{\hspace{1\tabcolsep}}C{1.05cm}@{\hspace{1\tabcolsep}}C{1.05cm}@{\hspace{1\tabcolsep}}C{1.15cm}@{\hspace{1\tabcolsep}}C{1.05cm}@{\hspace{1\tabcolsep}}C{1.05cm}}
				& $J^{*}$ & CPU & $\text{ESS}_{\text{MCMC}}$ & $\text{ESS}_{J^*}$ & \multicolumn{3}{c}{LPML ($10^3$)} & \multicolumn{3}{c}{$\text{ERR}_{\text{Mean}}$} & \multicolumn{3}{c}{$\text{ERR}_{\text{Dens}}$} \\
				& & & & & $Z_{1}$ & $Z_{2}$ & $Z_{3}$ & $Z_{1}$ & $Z_{2}$ & $Z_{3}$ & $Z_{1}$ & $Z_{2}$ & $Z_{3}$ \\\hline
				(i) & 23 & 10.24 & 619.1 & 1,352 & -0.93 & -0.74 & -0.35 & 2.60 & 3.85 & 9.41 & 80.83 & 105.49 & 8.28 \\
				(ii) & 23 & 12.15 & 516.0 & 1,117.2 & -1.75 & -1.4 & -0.73 & 2.60 & 3.86 & 9.71 & 87.85 & 103.60 & 9.32 \\ 
				(iii) & 23 & 4.42 & 560.5 & 1,070.9 & -0.88 & -0.96 & -0.34 & 2.33 & 2.13 & 8.59 & 75.34 & 79.74 & 7.66 \\ \hline
			\end{tabular}
		}
	}
	\caption[]{Simulated data set. Alternative scenarios: (i) longer MCMC chain; (ii) $n = 1,400$; (iii) no censoring. Summaries of the performance: computational burden,  mixing, goodness of fit, and predictive errors in mean and density obtained with the algorithm initialized at $J_0 = 15$.}
	\label{tab:Errors_grid_Simul_2}
\end{table}

\section{Application: life patterns of Colombian women}\label{app:application}

\begin{figure}[ht!]
	\begin{center}
		\includegraphics[width=0.5\textwidth]{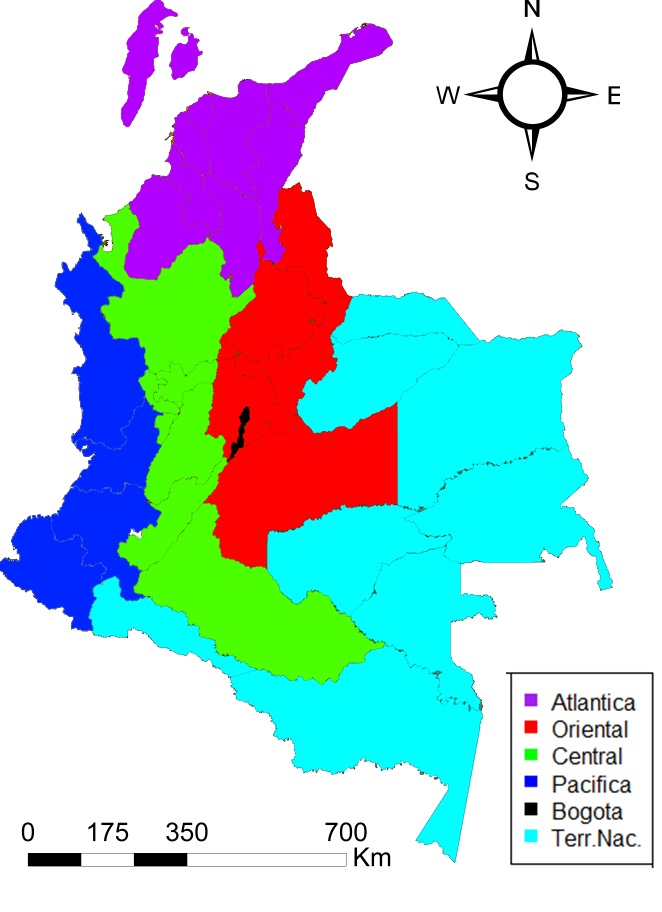} 
		\caption[]{Map of Colombia identifying the six regions considered by the DHS for the data collection and final report \citep{Profamilia_2011_DHS}, and adopted in this work.}
		\label{mapcolombia}
	\end{center}
\end{figure}

In the following sections, we provide some further discussion on the DHS data characteristics and report some additional figures, complementing and enriching the results reported in the main paper. We begin by providing a map of the regions of Colombia used in the study in Figure \ref{mapcolombia}. Different territorial divisions of Colombia are used in different contexts. We follow \cite{Profamilia_2011_DHS}, considering six regions: Atlantica, Oriental, Central, Pacifica, Bogota, Territorios Nacionales. We point out that Bogota is in fact a part of the Oriental region but was treated separately because of its peculiar features in terms of social and economic development. Territorios Nacionales could be divided into smaller, more homogeneous regions (e.g., Orinoquia and Amazon), but this is the definition employed by the DHS for the 2010 survey.

\subsection{The data: some considerations on sample weights}\label{sec:data}
The data arise from a complex survey design, and thus have associated weights with a rather complicated structure due both to survey design and to additional post-stratification carried out to adjust for various factors \citep[e.g., the total number of women interviewed in each municipality, non-response, etc.; see][for details]{Profamilia_2011_DHS}. This creates difficulties in further adjusting the weights based on the filtering considered (see Section 2 in the main text). Specifically, we do not know the population proportions for each filter within each municipality (e.g. of ethnicity, or of residents that have lived in the same region from at least the age of 6). Thus, the weights for the filtered data would need to be post-adjusted based on the observed sample proportions, leading to uncertainty in the weights. In addition, how this correction should relate to the post-stratification is unclear. We emphasize that even if we were able to obtain accurate weights for the filtered data, the use of survey weights in regression analyses is debated within the literature. From a frequentist perspective, \cite{winship1994} point out that when ordinary least squares (OLS) and weighted OLS lead to similar estimates the former is preferred, because it leads to improved efficiency, power, coverage. In fact, the authors state that ``where the sampling weights are solely a function of the independent variables" (as in our case), ``unweighted OLS estimates are preferred because they are unbiased, consistent, and have smaller standard errors." Moreover, the authors suggest that when OLS and weighted OLS estimates are substantially different one should investigate more flexible models, with e.g. non-linearity or interactions. Thus, since our model is nonparametric and can flexibly recover non-linearity, interactions, multi-modality (that may be due to missing or unmeasured covariates), we prefer the unweighted median regression and density regression estimates. 
 
From a Bayesian perspective, although our model and algorithm could be adjusted to include the sample weights, for example through a pseudo (i.e. weighted) likelihood or via data augmentation, as outlined by \cite{gunawan2017}, this may lead to Bayesian interval estimates that do not have the correct frequentist coverage. Indeed, the authors show that this is the case with the pseudo likelihood approach, and provide empirical evidence that data augmentation can perform better. In summary, due to the issues involved in the determination of accurate weights for the filtered data and to the poor coverage and efficiency of the weighted estimates that has been identified in literature, we carry out inference based on the unweighted data, which is further supported by the flexibility of our model.

\subsection{Additional figures}  
\begin{figure}[h]
		\includegraphics[width=0.95\textwidth]{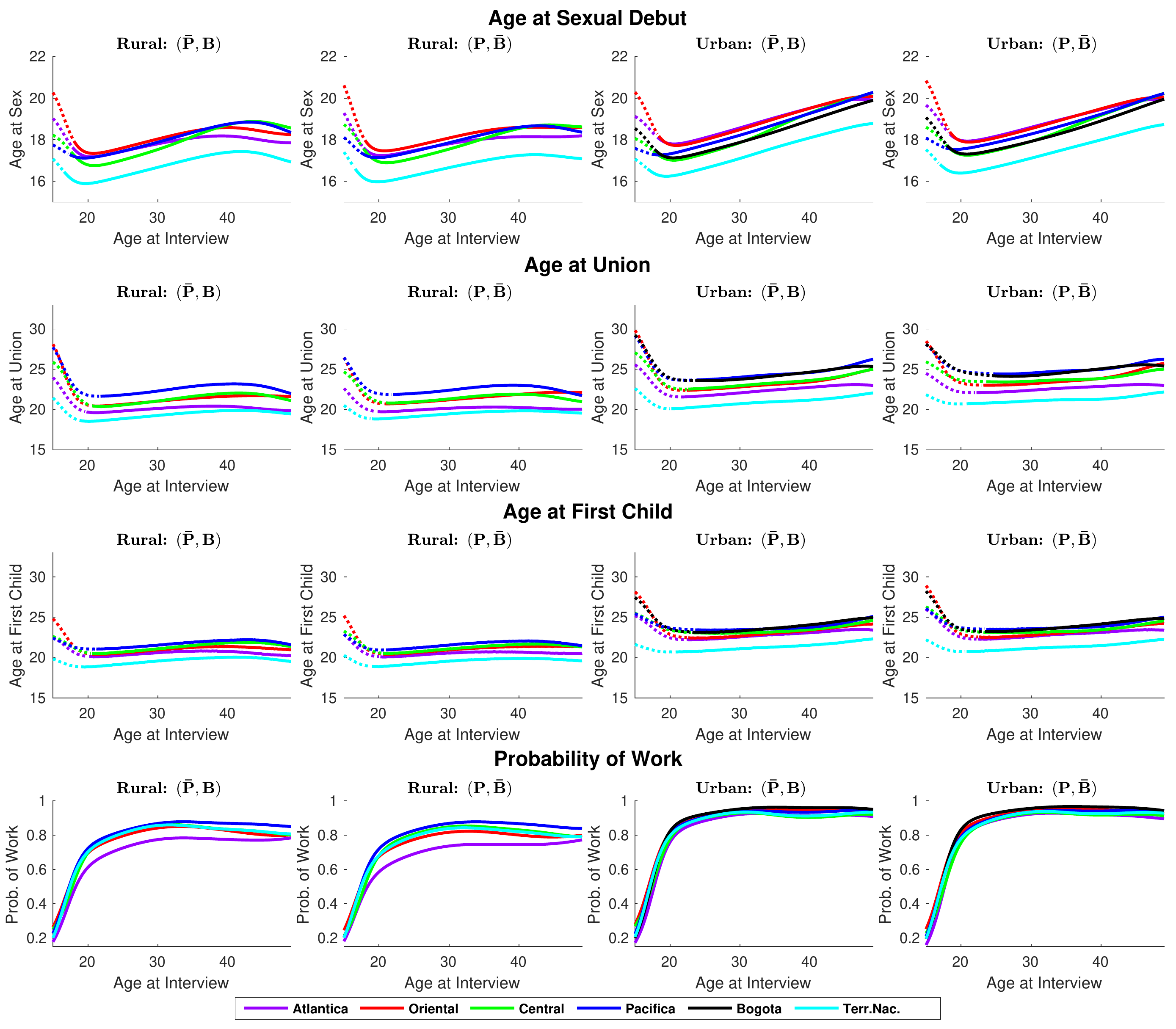}
		\caption{Predictive medians of the ages at sexual debut, union and child, and posterior probability of working, as functions of \textit{Age}, for women who grew up in violent environments with only physical punishment or only parental domestic violence, i.e (${\mathbf{P}}, {\mathbf{\bar{B}}}$) or (${\mathbf{\bar{P}}}, {\mathbf{B}}$). 
		Dotted lines indicate when the median exceeds \textit{Age}. Combined with Figure 5, observe that median ages increase as violence levels decrease, while the probability of working increases in younger cohorts for greater violence levels. This provides evidence for an anticipation of adulthood as violence levels increase.}
		\label{app_medwork_other}
\end{figure}

\begin{figure}[h]
		\includegraphics[width=0.95\textwidth]{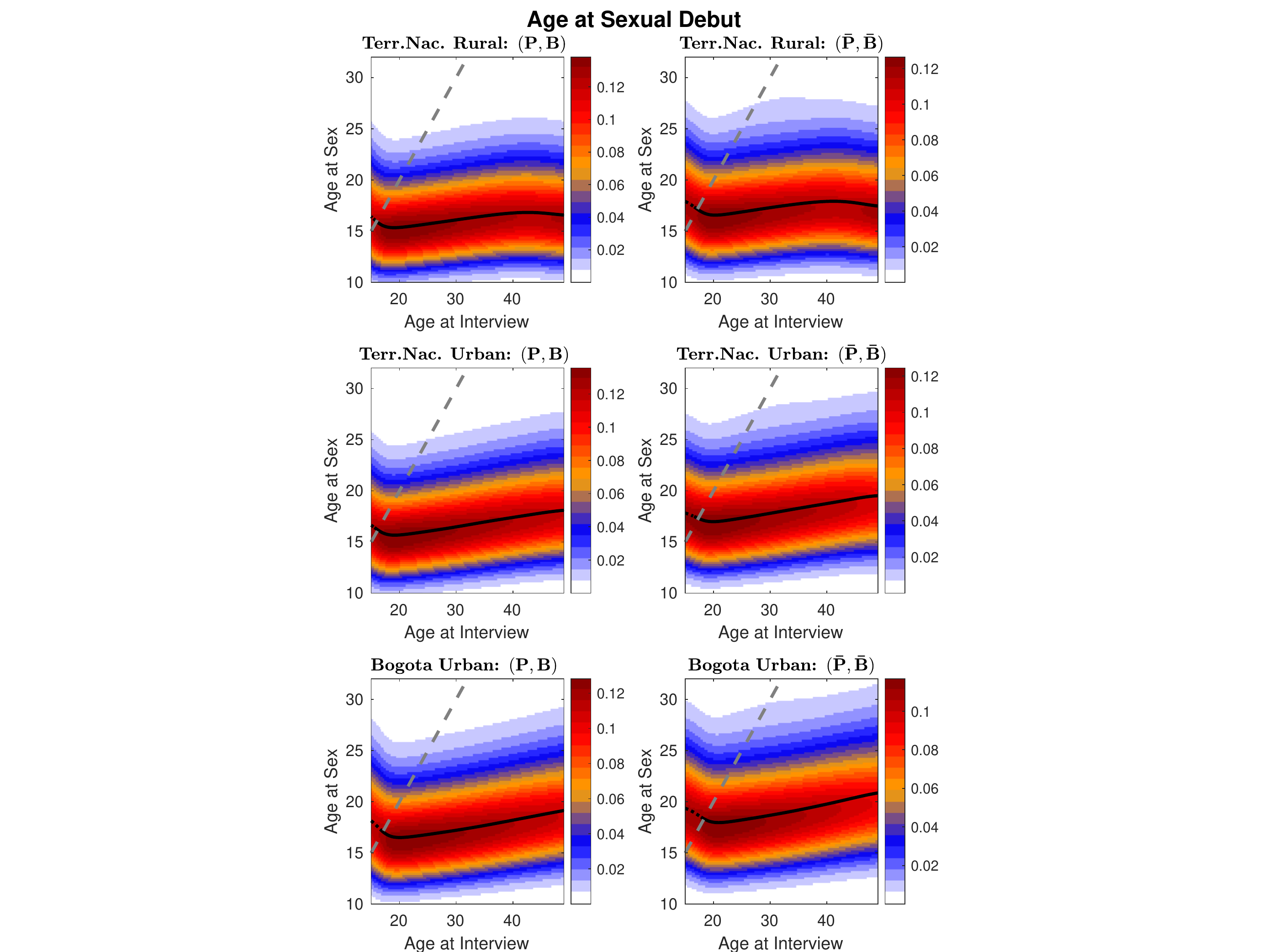}
		\caption{Predictive density of the age at sexual debut as a function of \textit{Age} for women who grew up in violent (${\mathbf{P}}, {\mathbf{B}}$) and non-violent families (${\mathbf{\bar{P}}}, {\mathbf{\bar{B}}}$). Analogously to Figure 6, results are reported for urban and rural areas of the least developed region (Territorios nacionales) and for the capital (Bogota). The region above the dashed line indicates when age at event exceeds \textit{Age}. The black line is the posterior median function. The median represents well the center of the distribution, and a decrease in both the median and dispersion of sexual debut is observed in younger cohorts, particularly in urban and developed regions.}
		\label{pred_densities_sex}
\end{figure}

\begin{figure}[h]
		\includegraphics[width=0.95\textwidth]{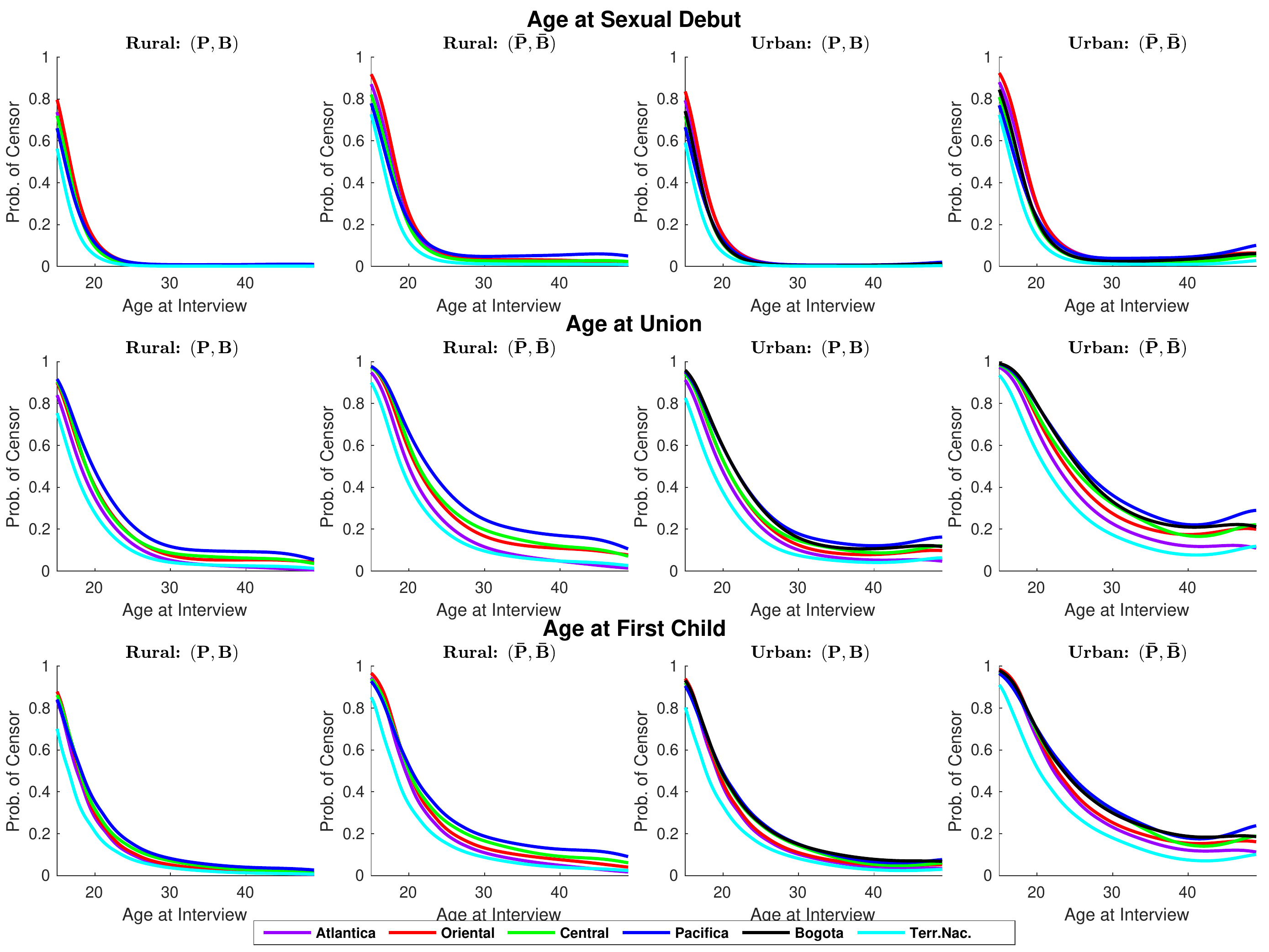}
		\caption{The predictive probability of censoring represents the probability that a woman will experience the event after the specified  \textit{Age} and is depicted for the events of sexual debut, union and child as a function of \textit{Age}, for women who grew up in violent (${\mathbf{P}}, {\mathbf{B}}$) and non-violent families (${\mathbf{\bar{P}}}, {\mathbf{\bar{B}}}$). Equivalently, the censoring probability represents the mass above the dashed line for a given \textit{Age} in the density plots of Figures 6 and \ref{pred_densities_sex}; when the right tail in the density exceeds the dashed line, interpreting the censoring probability is more reliable than focusing on the shape of the right tail. As expected, higher censoring probabilities are observed for younger cohorts and more developed regions and for the age at union and child over sexual debut. }
		\label{censorprob}
\end{figure}

\begin{figure}[h]
		\includegraphics[width=0.95\textwidth]{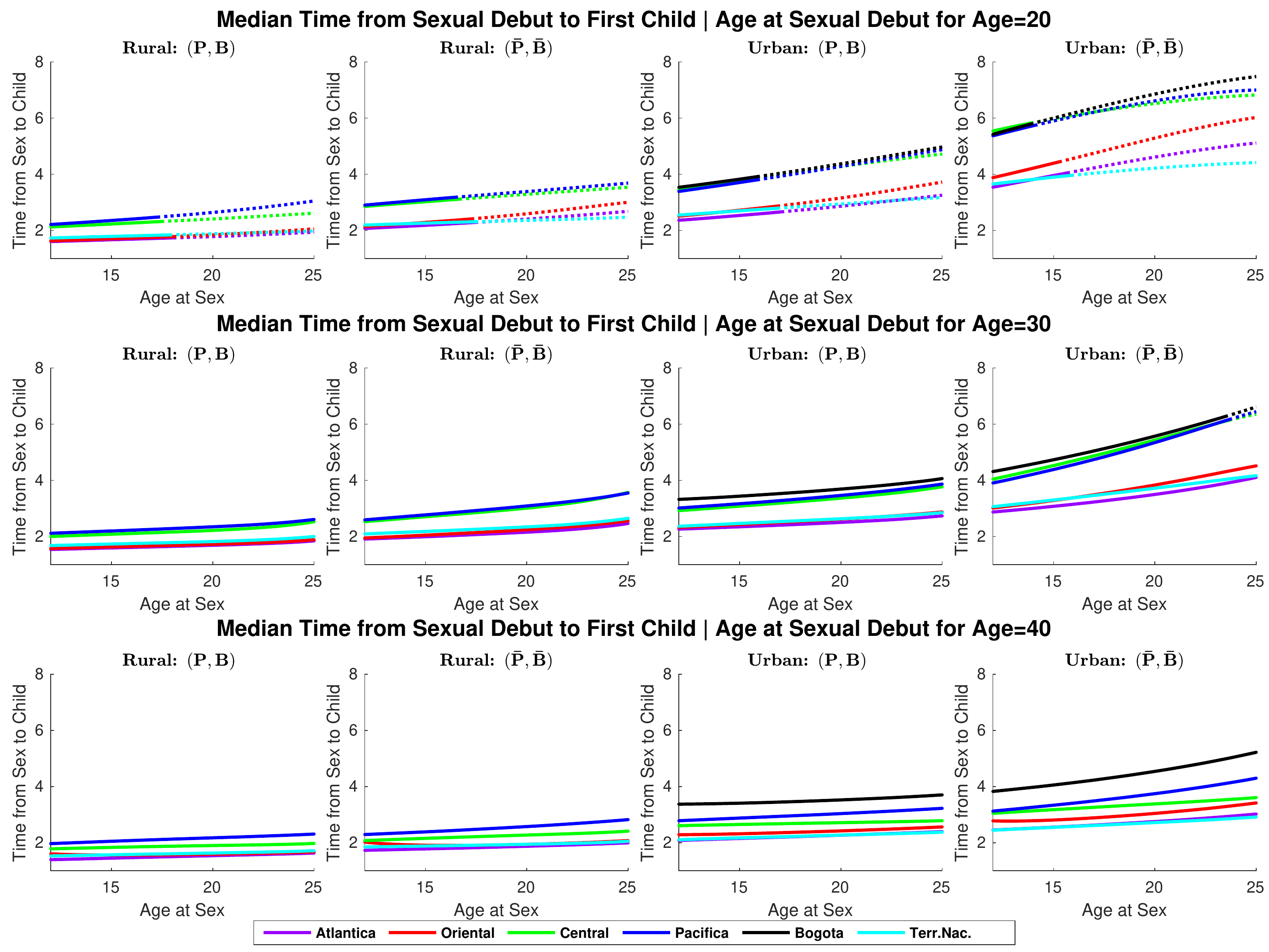}
		\caption{Conditional predictive medians of the time from sexual debut to first child given the age at sexual debut, as a function of the latter, for women  with $\textit{Age}=20, 30, 40$,  who grew up in violent (${\mathbf{P}}, {\mathbf{B}}$) and non-violent families (${\mathbf{\bar{P}}}, {\mathbf{\bar{B}}}$). Dotted lines indicate when the age at child is higher than the \textit{Age}. Notice that medians are higher for younger cohorts; thus, although we observe an anticipation of sexual debut in younger generations in Figure 5, these women tend to wait longer between sexual debut and first child. We can also appreciate a polarization between Atlantica, Oriental, and Territorios Nacionales on one side and Central, Pacifica, and Bogota on the other, particularly as \textit{Age} increases.}
		\label{app_medians1_conditional}
\end{figure}

\begin{figure}[h]
		\includegraphics[width=0.95\textwidth]{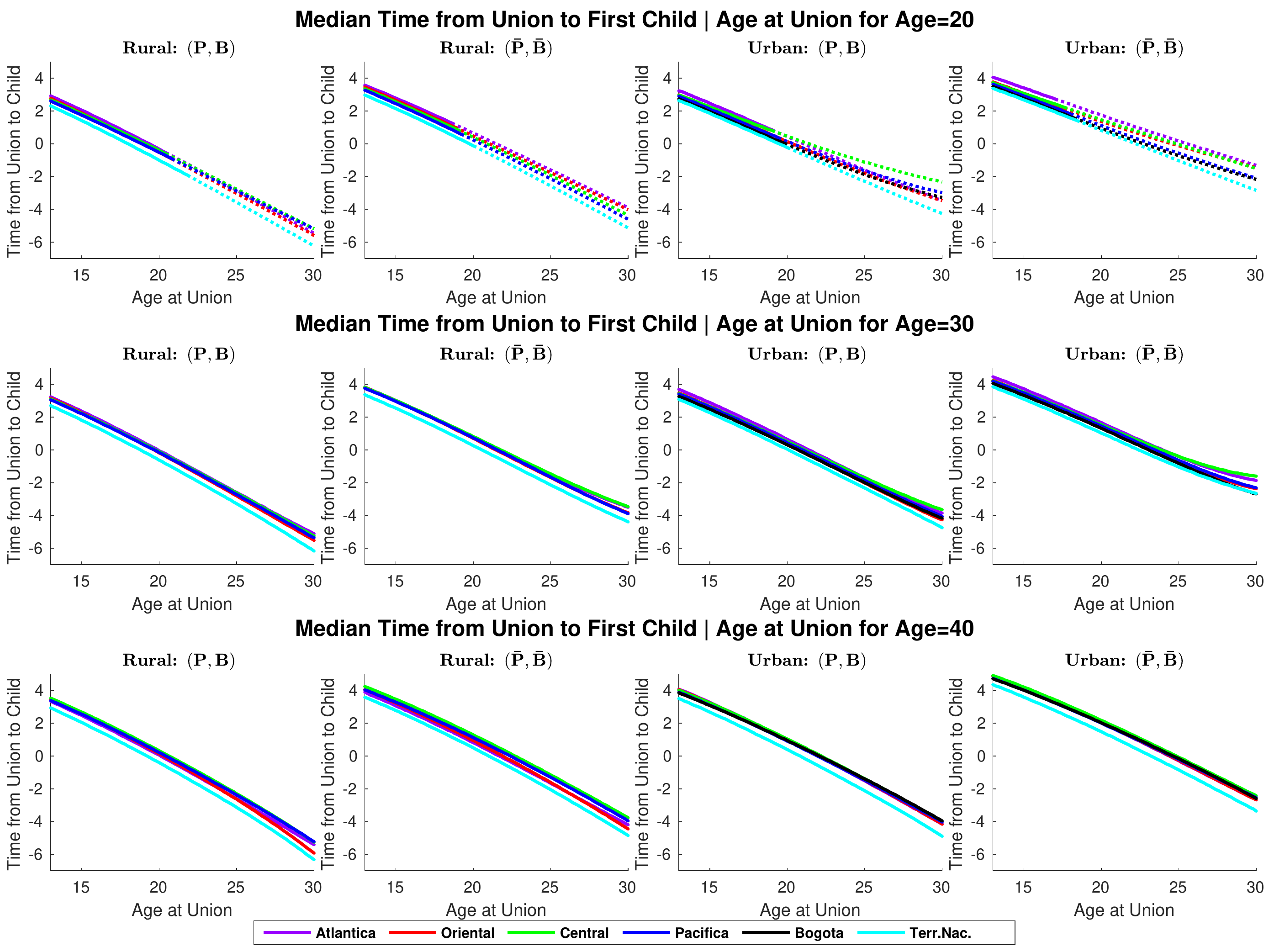}
		\caption{Conditional predictive medians of the time from union to first child given the age at union, as a function of the latter, for women aged 20, 30, and 40 at interview and who grew up in violent (${\mathbf{P}}, {\mathbf{B}}$) and non-violent families (${\mathbf{\bar{P}}}, {\mathbf{\bar{B}}}$). Dotted lines indicate when the age at child is higher than the \textit{Age}. As can be expected, median time from union to child decreases with age at union. Indeed, it is negative for high values of age at union, particularly in rural areas and for violent family environments, suggesting a greater tendency to have children out of wedlock.}
		\label{app_medians2_conditional}
\end{figure}

\begin{figure}[h]
	\includegraphics[width=0.95\textwidth]{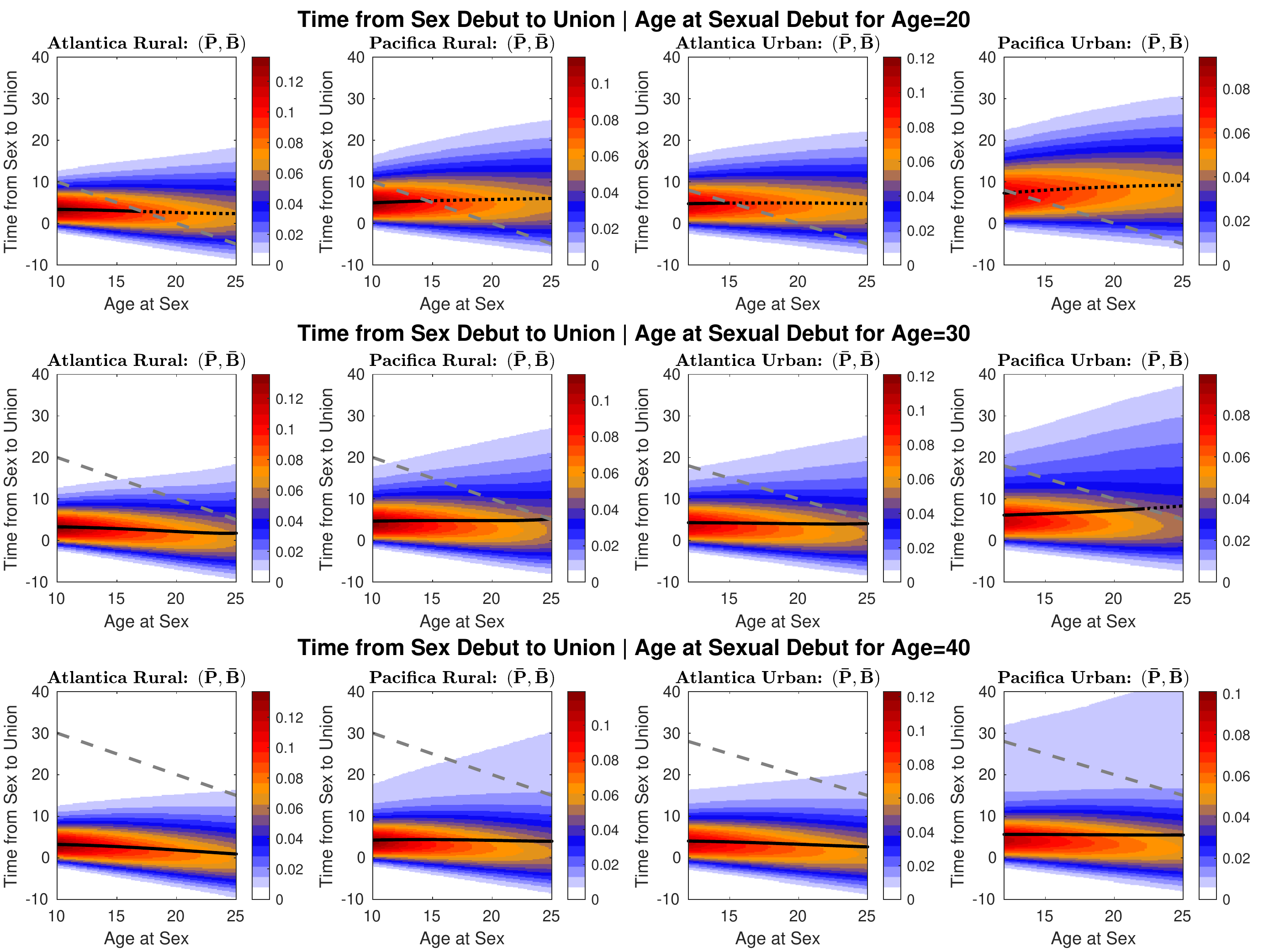}
		\caption{Conditional predictive density of the time from sexual debut to union given age at sexual debut, as a function of the latter, for women with $\textit{Age}=20, 30, 40$. Results are shown for women who grew up in a non-violent family (${\mathbf{\bar{P}}}, {\mathbf{\bar{B}}}$) and for urban and rural areas of Atlantic and Pacifica. The region above the dashed line indicates when age at union exceeds \textit{Age}. Combined with Figure 7, we observe that women in Pacifica and Bogota compared with Atlantica and Territorios Nacionales (and to a lesser extent Oriental) not only have a higher median time from sexual debut to union but also increased dispersion and a heavier right tail, reflecting a wider variety of choices for women to delay union after sexual debut in these regions. Additionally, a slight increase in median time and dispersion can be appreciated for decreasing \textit{Age}, supporting a weaker relation between sexual debut and union in younger cohorts, that is more evident in developed urban areas.} 
		\label{app_density_conditional_sextounion}
\end{figure}

\begin{figure}[h]
		\includegraphics[width=0.95\textwidth]{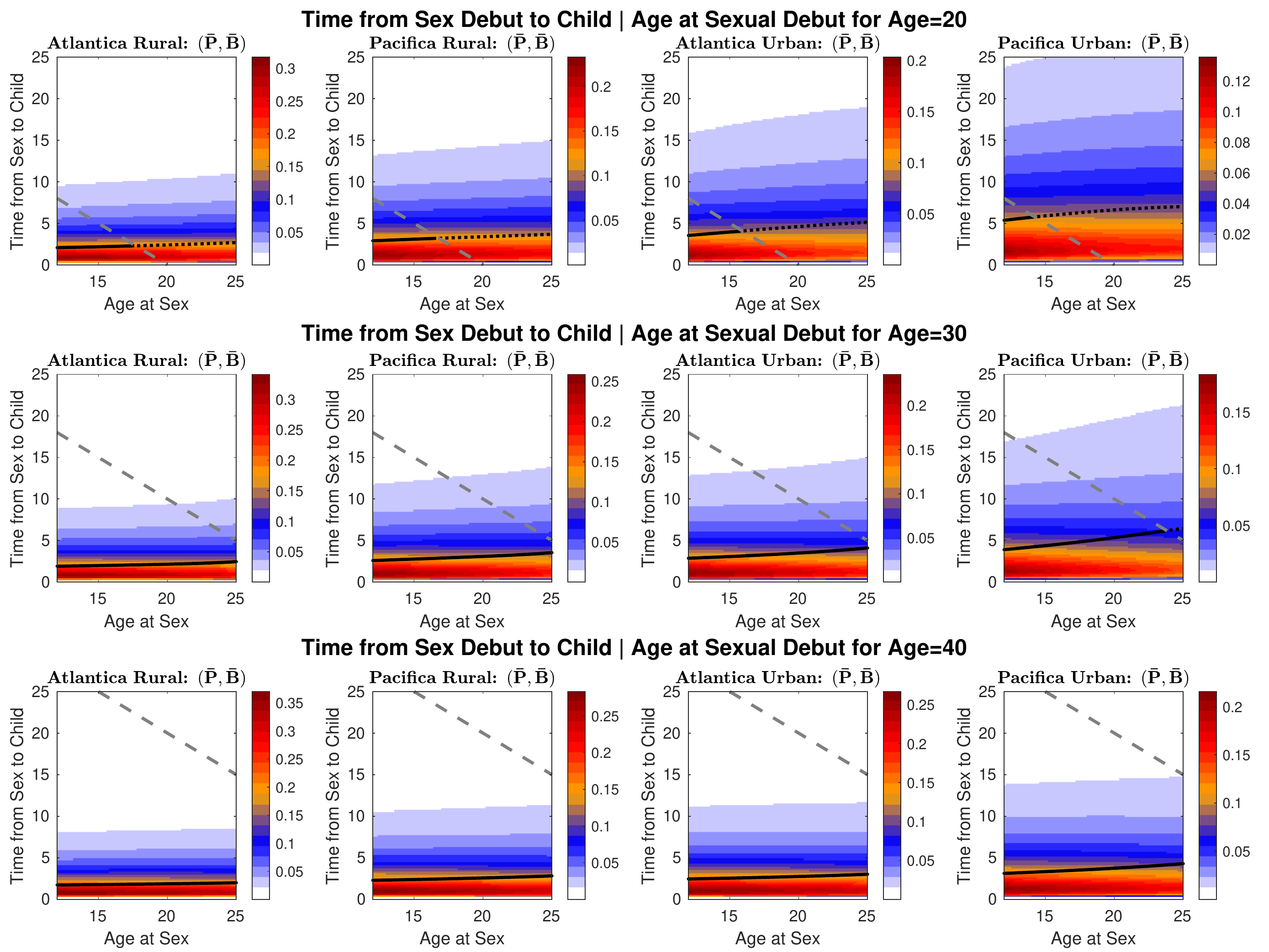}
		\caption{Conditional predictive density of the time from sexual debut to child given age at sexual debut, as a function of the latter, for women  with $\textit{Age}=20, 30, 40$. Results are shown for women who grew up in a non-violent family (${\mathbf{\bar{P}}}, {\mathbf{\bar{B}}}$) and for urban and rural areas of Atlantic and Pacifica. The region above the dashed line indicates when age at child exceeds \textit{Age}. The heavier right tail, reflecting a wider variety of choices for women to delay motherhood after sexual debut, is evident as \textit{Age} increases, particularly in developed urban areas. This supports the claim of a weaker relation between sexual debut and motherhood in younger cohorts. }
		\label{app_density_conditional_sextochild}
\end{figure}

\begin{figure}[h]
	\includegraphics[width=0.95\textwidth]{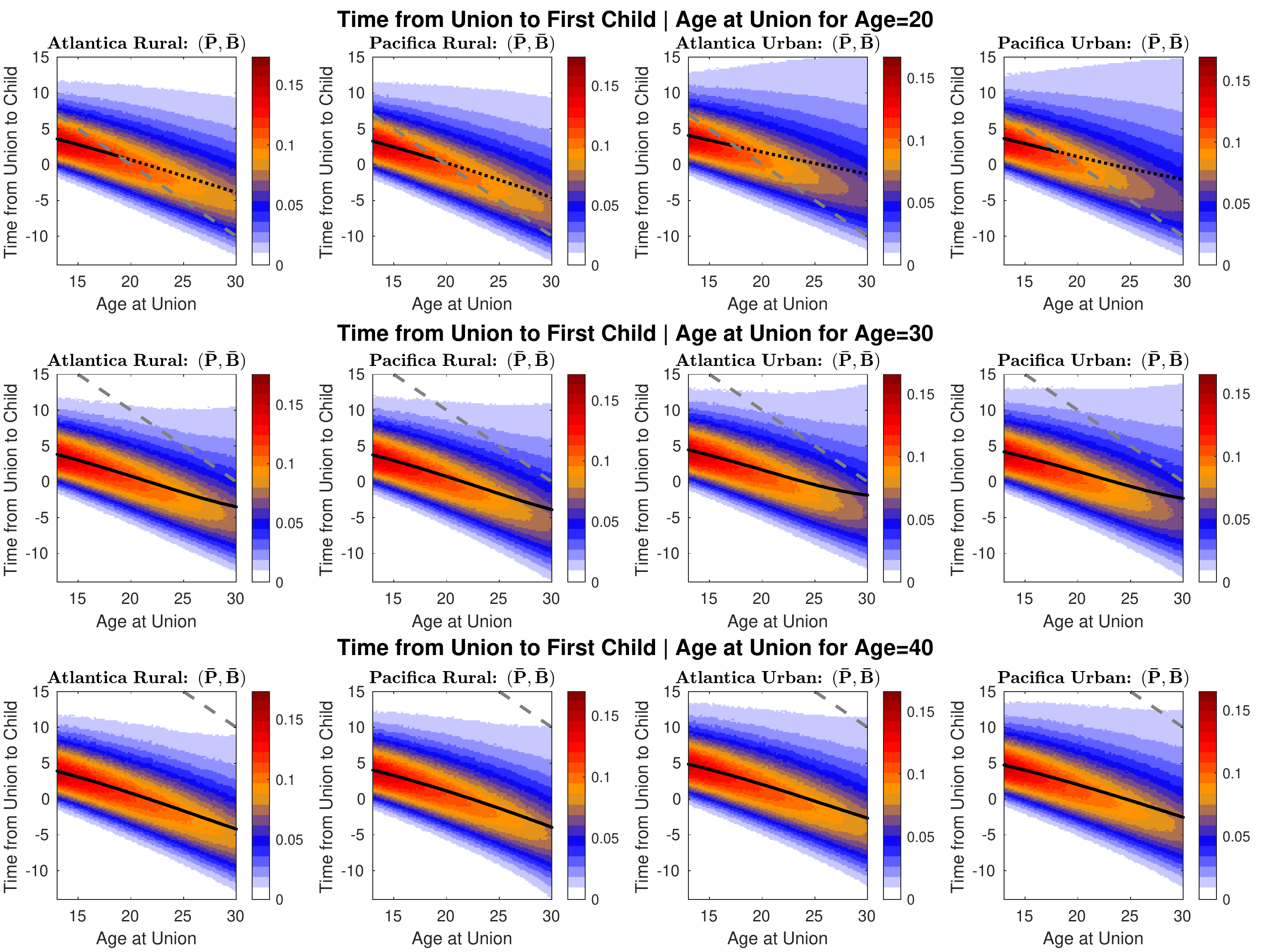}
		\caption{Conditional predictive density of the time from union to first child given age at union, as a function of the latter, for women with $\textit{Age}=20, 30, 40$. Results are shown for women who grew up in a non-violent family (${\mathbf{\bar{P}}}, {\mathbf{\bar{B}}}$) and for urban and rural areas of Atlantic and Pacifica. The region above the dashed line indicates when age at first child exceeds \textit{Age}.} 
		\label{app_density_uniontochild}
\end{figure}

\begin{figure}[h]
	\includegraphics[width=0.95\textwidth]{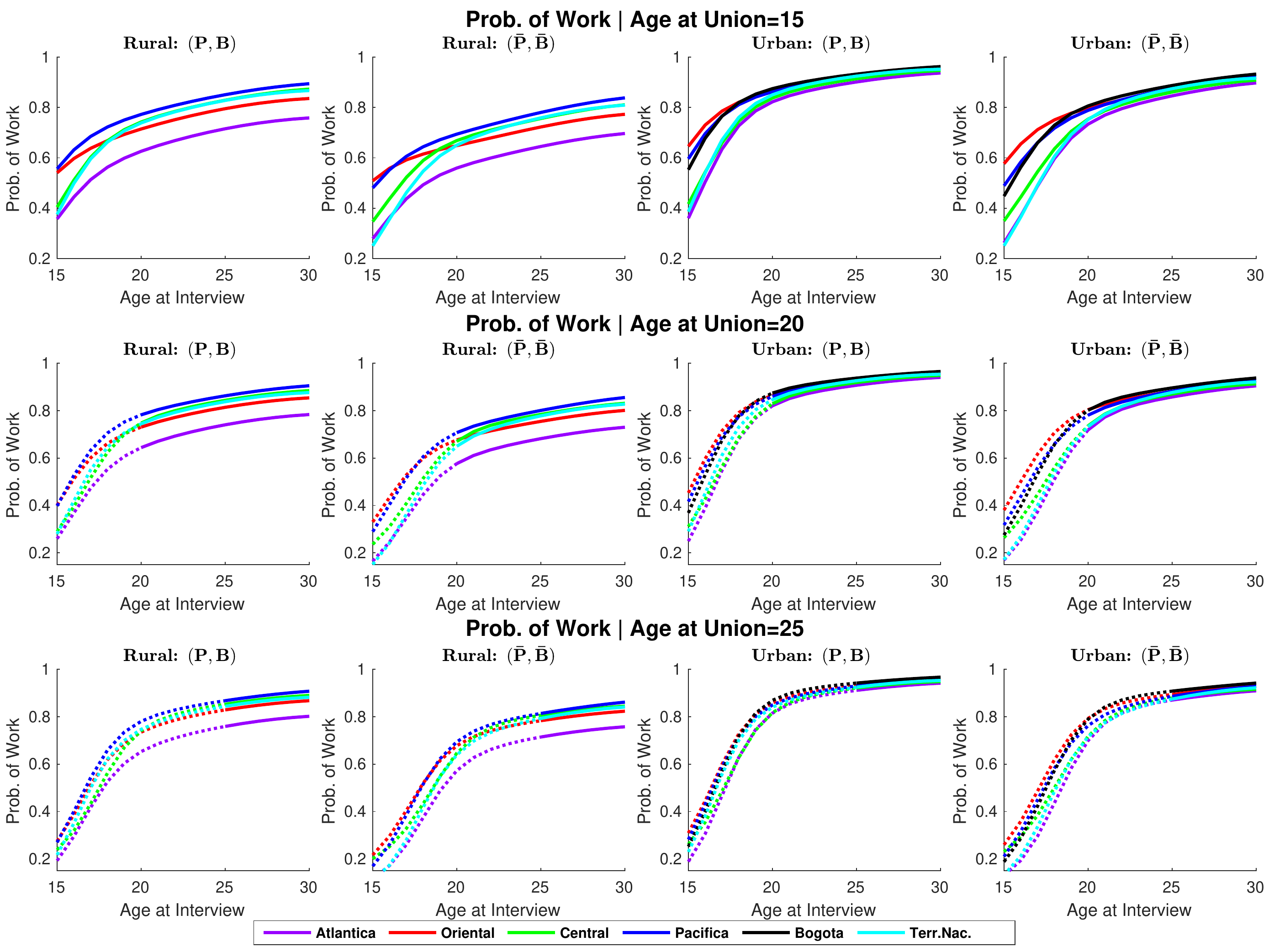}
		\caption{Conditional predictive probability of working as function of $Age$ given different ages at union, for women who grew up in violent (${\mathbf{P}}, {\mathbf{B}}$) and non-violent families (${\mathbf{\bar{P}}}, {\mathbf{\bar{B}}}$). 
			Dotted lines indicate when \textit{Age} is less than the age at event. While we observe an increased probability of working for young cohorts that established an early union, in contrast to Figure 8, no scaring effect is visible, i.e. the probability of working in older cohorts is unaffected by the conditioned age at union. }
		\label{app_4_work_conditional}
\end{figure}

We display additional figures, enriching the results reported in the main text, and for convenience, comments on possibly relevant findings are reported in the figures' captions. 
Figure \ref{app_medwork_other} complements Figure 5 by reporting median ages at events for women who grew up in violent environments with only physical punishment or only parental domestic violence, i.e (${\mathbf{P}}, {\mathbf{\bar{B}}}$) or (${\mathbf{\bar{P}}}, {\mathbf{B}}$).  
Figure \ref{pred_densities_sex} completes Figure  6 by displaying the predictive density of the age at sexual debut. 
Figure \ref{censorprob} reports the predictive probability of censoring, that is the probability that a woman will experience the event after  the given  \textit{Age}, as a function of \textit{Age}. 
Turning to the conditional analysis, the conditional predictive medians for the time from sexual debut to first child given the age at sexual debut is shown in Figure \ref{app_medians1_conditional} and for the time from union to first child given the age at union is shown in Figure \ref{app_medians2_conditional}. The underlying conditional predictive densities for selected covariate combinations are visualized in Figures \ref{app_density_conditional_sextounion}, \ref{app_density_conditional_sextochild}, and \ref{app_density_uniontochild}. 
Finally, to explore the possible relation between anticipation of union on work activity, Figure	\ref{app_4_work_conditional} reports the conditional predictive probability of working as function of $Age$ given different ages at union.


\clearpage
\bibliographystyle{ba}
\bibliography{ref_Combined}
